\documentclass[sigconf]{acmart}
\AtBeginDocument{%
 }
   
\usepackage{adjustbox,bbm,multirow,multicol}
\usepackage{natbib}
\usepackage{enumitem} 
\usepackage[normalem]{ulem}
\usepackage{algorithm}
\usepackage{algorithmic}
\usepackage{bbding} 
\usepackage{makecell}
\PassOptionsToPackage{table}{xcolor} 
\usepackage{colortbl}  
\usepackage{nicematrix, tikz}
\setcopyright{acmcopyright}
\acmDOI{XXXXXXX.XXXXXXX}

\acmPrice{15.00}
\acmISBN{}

\begin{document}

\title[\textit{SAKE}: Self-aware Knowledge Exploitation-Exploration for GMNER]{\textit{SAKE}: Self-aware Knowledge Exploitation-Exploration for Grounded Multimodal Named Entity Recognition}

\author{Jielong Tang}
\affiliation{%
  \institution{Sun Yat-sen University}
  \country{Zhuhai, China}}
\email{tangjlong3@mail2.sysu.edu.cn}

\author{Xujie Yuan}
\affiliation{%
  \institution{Sun Yat-sen University}
  \country{Zhuhai, China}}
\email{yuanxj8@mail2.sysu.edu.cn}

\author{Jiayang Liu}
\affiliation{%
  \institution{Sun Yat-sen University}
  \country{Zhuhai, China}}
\email{liujy639@mail2.sysu.edu.cn}

\author{Jianxing Yu}
\affiliation{%
  \institution{Sun Yat-sen University}
  \country{Zhuhai, China}}
\email{yujx26@mail.sysu.edu.cn}

\author{Xiao Dong}
\affiliation{%
  \institution{Shandong Normal University}
  \country{Jinan, China}}
\email{dx.icandoit@gmail.com}

\author{Lin Chen}
\affiliation{%
  \institution{University of the Chinese Academy of Sciences}
  \country{Beijing, China}}
\email{chenlin@iie.ac.cn}

\author{Yunlai Teng}
\affiliation{%
  \institution{China Mobile Group}
  \country{Beijing, China}}
\email{tengyunlaiit@chinamobile.com}

\author{Shimin Di}
\affiliation{%
  \institution{Southeast University}
  \country{Nanjing, China}}
\email{shimin.di@seu.edu.cn}

\author{Jian Yin}
\authornote{Jian Yin is the corresponding author.}
\affiliation{%
  \institution{Sun Yat-sen University}
  \country{Zhuhai, China}}
\email{issjyin@mail.sysu.edu.cn}

\begin{abstract}
Grounded Multimodal Named Entity Recognition (GMNER) aims to extract named entities and localize their visual regions within image–text pairs, serving as a pivotal capability for various downstream applications. In open-world social media platforms, GMNER remains challenging due to the prevalence of long-tailed, rapidly evolving, and unseen entities. To tackle this, existing approaches typically rely on either external knowledge exploration through heuristic retrieval or internal knowledge exploitation via iterative refinement in Multimodal Large Language Models (MLLMs). However, heuristic retrieval often introduces noisy or conflicting evidence that degrades precision on known entities, while solely internal exploitation is constrained by the knowledge boundaries of MLLMs and prone to hallucinations. The core limitation lies in the fact that current models lack awareness of their own knowledge boundaries, and thus cannot adaptively think when to rely on internal memory versus when to seek external evidence. To address this, we propose \textbf{SAKE}, an end-to-end agentic framework that harmonizes internal knowledge exploitation and external knowledge exploration via self-aware reasoning and adaptive search tool invocation. We implement this via a two-stage training paradigm. First, we propose Difficulty-aware Search Tag Generation, which quantifies the model's entity-level uncertainty through multiple forward samplings to produce explicit knowledge-gap signals. Based on these signals, we construct \textbf{SAKE-SeCoT}, a high-quality Chain-of-Thought dataset that equips the model with basic self-awareness and tool-use capabilities through supervised fine-tuning. Second, we employ agentic reinforcement learning with a hybrid reward function that penalizes unnecessary retrieval, enabling the model to evolve from rigid search imitation to genuine self-aware decision-making about when retrieval is truly necessary. Extensive experiments on two widely used social media benchmarks demonstrate that SAKE achieves state-of-the-art performance, outperforming the previous best method by 3.75\% and 2.91\% on GMNER F1, while reducing the average search rate to 68.8\%. Our code is released at \url{https://github.com/tangjielong928/SAKE}.
\end{abstract}






\maketitle

\section{Introduction}

\begin{figure}[!t]
	\centering
	\includegraphics[width=\linewidth]{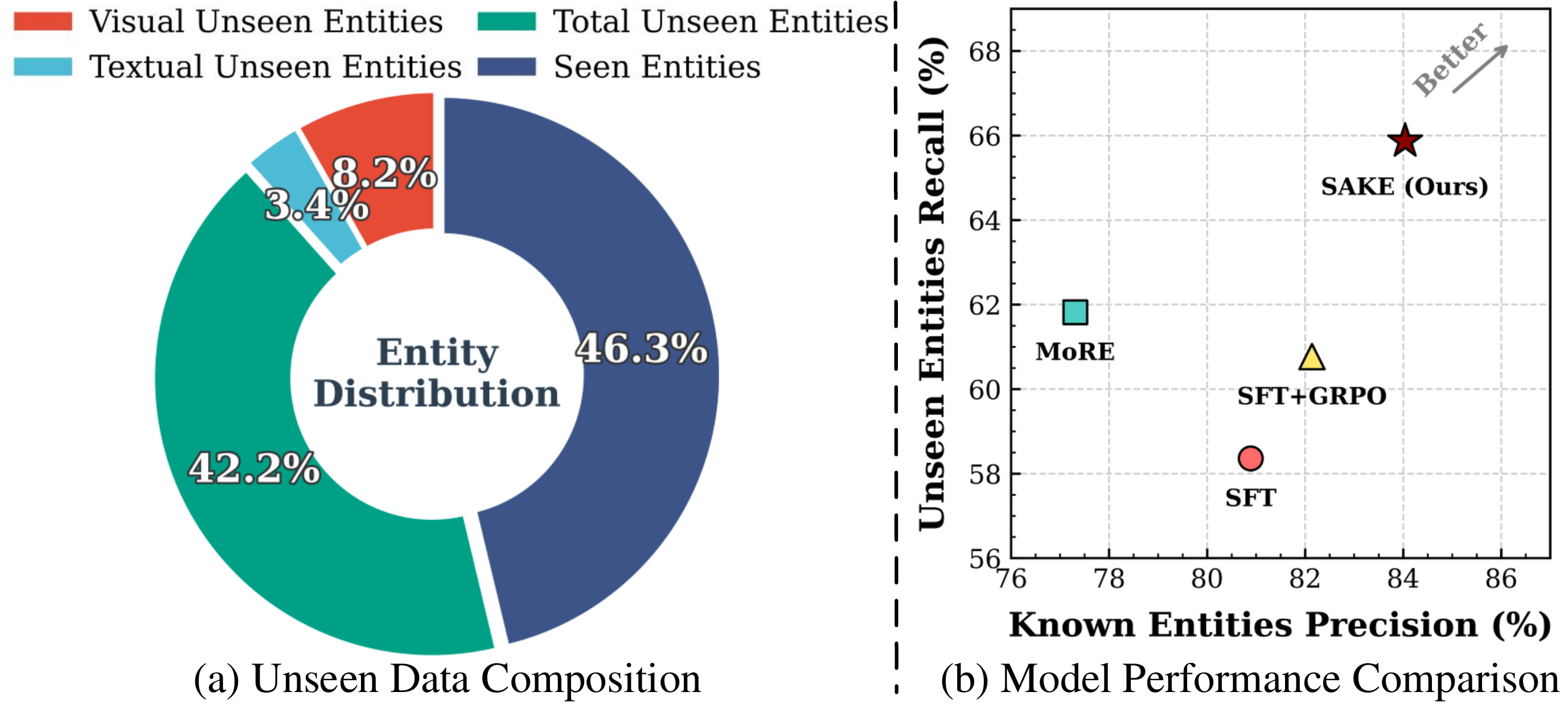}
	\caption{(a) Unseen data composition of the Twitter benchmark~\cite{gmner}. (b) Performance comparison of SAKE against different training paradigms based on Qwen2.5-VL-7B-Instruct.}
	\label{intro_data_analysis}
\end{figure}

Named Entity Recognition (NER) aims to extract entity mentions and their types from text, while Multimodal NER (MNER) leverages the paired image as auxiliary information. Grounded Multimodal Named Entity Recognition (GMNER) goes one step further, grounding each entity to its corresponding visual region in the image. Such entity-level grounding underpins a wide range of downstream applications, including multimodal knowledge graph construction~\cite{mmkg1} and knowledge-intensive visual question answering (VQA)~\cite{li2025answering}.
Despite its significance, GMNER in open-world social media platforms remains highly challenging, as multimodal entities typically follow long-tailed distributions, undergo rapid evolution, and exhibit open-world characteristics. As shown in 
Figure~\ref{intro_data_analysis}(a), \textit{unseen entities}\footnote{In this work, unseen refers to the absence of one of the entity's multimodal attributes (text mention, type, or visual region) in the training set.} account for a dominant proportion ($\approx 54\%$) of widely used Twitter benchmarks~\cite{gmner},
making it difficult for previous supervised fine-tuning (SFT) methods~\cite{wang2022ita, gmner, mqspn, fg-gmner} to 
generalize outside their training distribution.


\begin{figure}[]
\centering
\includegraphics[width=1.0\linewidth]{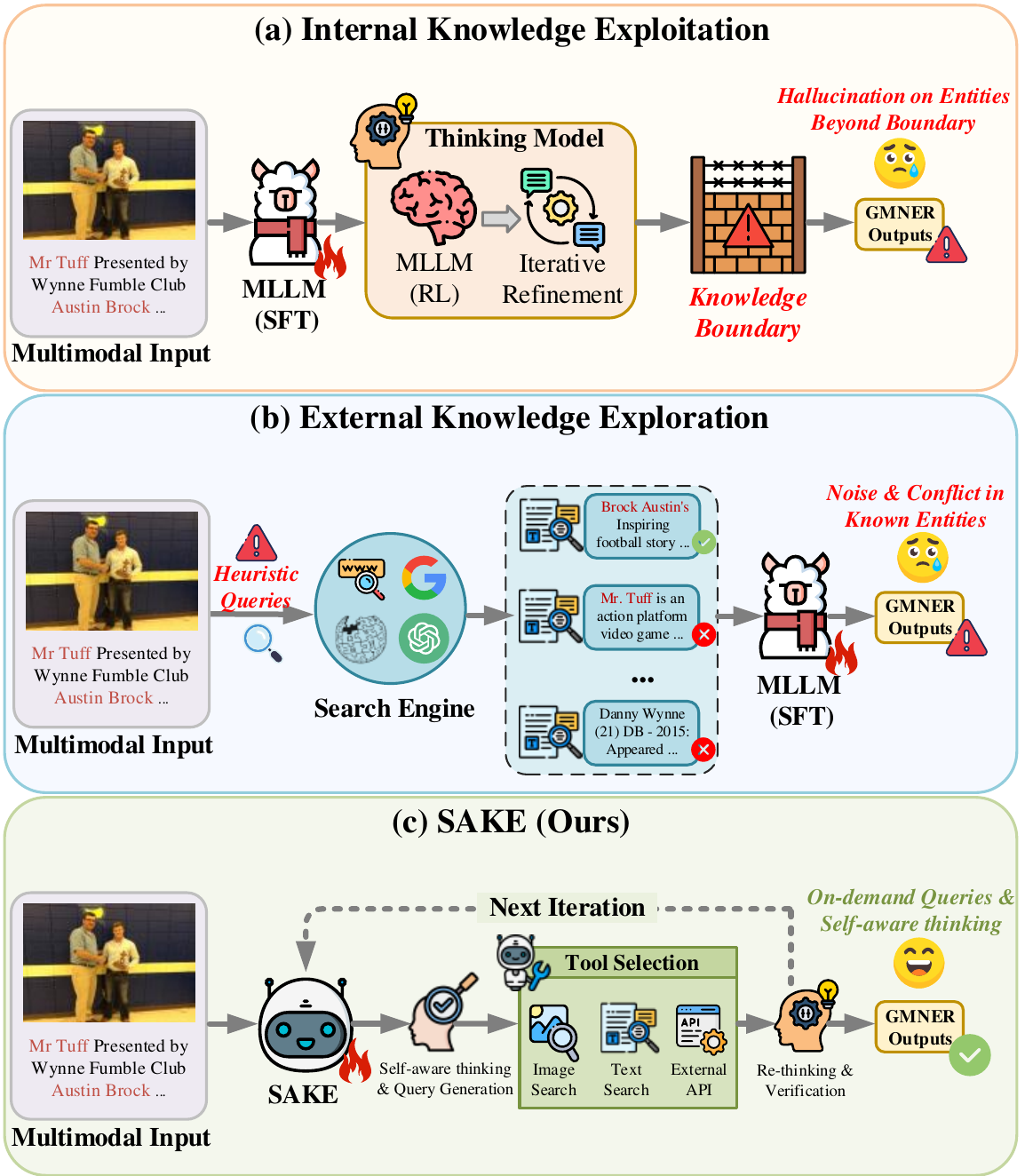}
\caption{Comparison of existing GMNER approaches. (a) Internal knowledge exploitation relies solely on parametric memory, leading to hallucinations when encountering entities beyond the knowledge boundary. (b) External knowledge exploration often introduces noisy or conflicting evidence. (c) SAKE adaptively invokes external tools for on-demand retrieval only when needed, bridging internal reasoning and external knowledge.}
\label{intro}
\end{figure}

Recent advances in multimodal large language models (MLLMs) offer opportunities to address this challenge through their vast pre-training knowledge and generalization. 
Building upon this capability, current state-of-the-art methods~\cite{yu2025isr, lin2025makar, tang2025-unco} employ reinforcement learning (RL)~\cite{guo2025deepseek} or prompt-based Chain-of-Thought (CoT)~\cite{wei2022chain} to elicit internal parametric knowledge, thereby facilitating the iterative refinement of unseen multimodal entities. We conceptualize this process as \textit{Internal Knowledge Exploitation}, as illustrated in Figure~\ref{intro}(a). However, relying solely on the internal knowledge encounters the \textit{knowledge boundary limitation}: when an MLLM is confronted with multimodal entity attributes it has never internalized (e.g., the specific face or attire of a niche influencer), it lacks adequate awareness of this knowledge gap, and produces plausible but incorrect recognition. 

To break through the knowledge boundary, some works attempt to \textit{explore} external knowledge sources. MoRE~\cite{wang2022named} retrieves external knowledge for all entities through multi-modal search, while other works~\cite{ok2024scanner, lin2025makar} employ similar strategies using heuristic queries (e.g., original images or phrases). This \textit{External Knowledge Exploration} strategy can enrich entity-related knowledge beyond parametric memory.
However, indiscriminate exploration is not only computationally expensive but also inevitably introduces noisy and irrelevant information. For instance, as shown in Figure~\ref{intro}(b), the retrieved results for the entity \textit{``Mr~Tuff''} indicate a video game, which is inconsistent with the \textit{``Person''} entity depicted in the image. This external noise directly conflicts with the model's correct internal belief that \textit{``Mr~Tuff''} refers to a person.
This raises a problem: aggressive exploration improves unseen entity recall but degrades known entity precision through noise injection. As evidenced in Figure~\ref{intro_data_analysis}(b), MoRE's indiscriminate exploration policy achieves 61.8\% recall on unseen entities but suffers a 3.6\% precision drop on known entities compared to pure SFT methods (77.3\% vs 80.9\%).

Given these observations, we argue that the core of the current dilemma lies in two aspects: 
(1) \textit{Exploitation} of internal knowledge activates MLLM's parametric memory to improve extraction performance, but fails to handle concepts beyond the knowledge boundary; 
(2) \textit{Exploration} of external knowledge is often necessary for entities outside the knowledge boundary, yet can be detrimental when indiscriminately applied to known concepts, while also incurring high computational costs.
Existing methods typically rely on fixed policies (either exploiting internal knowledge or exploring external sources), lacking a mechanism to adaptively decide when to rely on internal memory or resort to external retrieval based on entity-specific contexts.

To bridge this research gap, we propose \textbf{SAKE} (\textit{Self-aware Knowledge  Exploitation-Exploration}), an end-to-end agentic framework that harmonizes knowledge internal exploitation and external exploration via self-aware reasoning. Specifically, SAKE operates through multi-turn interactions with external search tools (see Figure~\ref{intro}(c)). It first explicitly expresses its uncertainty over candidate entity concepts, and dynamically decides whether to directly answer (for certain entities) or to formulate on-demand queries (for uncertain concepts). Retrieved evidence is iteratively incorporated into the reasoning process, enabling the model to progressively resolve conceptual ambiguities and consolidate the final grounding. 
A key challenge in realizing self-awareness is that current MLLMs lack explicit signals of their knowledge gaps. Our preliminary experiments reveal that the base model (Qwen2.5-VL-7B) exhibits overconfidence in real-world entities when relying solely on prompt-based self-reflection~\cite{madaan2023self, ji2023towards}, resulting in extremely low tool-invocation rates (see Section~\ref{search_tag}).
To address this, we propose Difficulty-aware Search Tag Generation, which explicitly quantifies the model's uncertainty over a specific entity concept through multiple forward samplings, and generates pseudo search tags as explicit ``I am not sure'' signals. We construct \textbf{SAKE-SeCoT}, a high-quality Chain-of-Thought dataset based on these tags, equipping the base model with the capability of tool invocation through SFT.
However, SFT alone mainly induces imitation of retrieval behaviors, rather than adaptively thinking when retrieval is truly necessary. 
Inspired by existing research~\cite{guo2025deepseek, shao2024deepseekmath} that RL enables models to genuinely learn how to reason, we employ RL with a hybrid reward that penalizes unnecessary retrieval, enabling SAKE to transition from rigid imitation to self-aware decision-making. This allows SAKE to strategically arbitrate between internal knowledge exploitation and external exploration.

Our contributions could be summarized as follows:
\begin{itemize}[leftmargin=1.5em]
\item We propose \textbf{SAKE}, an end-to-end agentic model for real-world GMNER, which adaptively balances internal knowledge exploitation and external knowledge exploration with self-aware thinking and on-demand tool invocations.
\item We construct \textbf{SAKE-SeCoT}, a high-quality Chain-of-Thought dataset based on difficulty-aware search tag generation, which equips the base model with explicit self-awareness and effective tool-usage capabilities.
\item Extensive experiments on two widely used GMNER benchmarks demonstrate that SAKE consistently achieves state-of-the-art performance, particularly improving generalization to real-world multimodal entities.
\end{itemize}


\section{Related Work}
\textbf{Grounded Multimodal Named Entity Recognition.}
Grounded Multimodal Named Entity Recognition (GMNER) aims to extract textual entities and ground them to corresponding visual regions. Early approaches primarily focused on leveraging off-the-shelf object detectors~\cite{girshick2015fast, zhang2021vinvl, liu2023groundingDINO} to propose candidate regions, followed by end-to-end modeling, including sequence labeling~\cite{umt}, sequence generation~\cite{gmner,fg-gmner,li2024generative}, and set prediction~\cite{mqspn}, to predict entity-type-region triplets. Despite effectiveness, these methods often suffer from limited background knowledge and poor generalization, particularly within open-world social media scenarios. To address this limitation, recent studies~\cite{liu2024multi, yuan2026visual} have explored incorporating external knowledge. For instance, MoRE~\cite{wang2022named} and SCANNER~\cite{ok2024scanner} utilize external search engines to acquire entity-related context. However, these methods typically rely on heuristic retrieval strategies; such indiscriminate retrieval often introduces irrelevant noise or conflicting information. On the other hand, some approaches~\cite{tang2025-unco, yu2025isr} attempt to exploit the iterative self-refinement capabilities of Multimodal LLMs for entity grounding. RiVEG~\cite{li2024llms} and GEM~\cite{wang2024granular} leverage closed-source LLMs to acquire auxiliary knowledge. Nevertheless, relying solely on the internal parametric knowledge of MLLMs is inherently restricted by knowledge boundaries. MAKAR~\cite{lin2025makar} introduces a multi-agent framework that integrates knowledge exploration, entity correction, and reasoning-based grounding. Distinct from existing methods, \textit{SAKE} possesses self-awareness to determine \textit{when} retrieval is necessary and \textit{what} content to retrieve, effectively harmonizing external retrieved knowledge with internal parametric knowledge for GMNER. 

\textbf{Agentic Reinforcement Learning.} The integration of end-to-end reinforcement learning (RL) into post-training processes has accelerated the development of Large Reasoning Models (LRMs), as demonstrated by OpenAI’s o-series~\cite{jaech2024openai}, DeepSeek-R1~\cite{guo2025deepseek}, and Kimi-K1.5~\cite{team2025kimi}. This enhancement in reasoning capabilities has been pivotal in the rise of Deep Research agents~\cite{zheng2025deepresearcher}, which employ RL to autonomously orchestrate open-ended information retrieval. In contrast to static agent workflows~\cite{nakano2021webgpt, luo2023sail}, these RL-based agents optimize long-horizon planning, enabling the synthesis of information from diverse sources. Frameworks such as Search-R1~\cite{jin2025search}, Search o1~\cite{li2025search}, and ReSearch~\cite{chen2025learning} adopt similar paradigms to improve multi-turn retrieval, thus democratizing autonomous knowledge discovery. Recent advancements, including MMSearch-R1~\cite{wu2025mmsearch} and DeepEyes~\cite{zheng2025deepeyes, hong2025deepeyesv2}, extend this capability to the visual question answering (QA) domain, fostering emergent behaviors such as ``thinking with images''. Unlike previous works that are confined to standard QA tasks, \textit{SAKE} leverages agentic RL to drive autonomous interactions with open-world multimodal environments, enabling fine-grained entity perception through analogical reasoning over retrieved multimodal contexts.

\begin{figure*}[]
\centering
\includegraphics[width=1.0\linewidth]{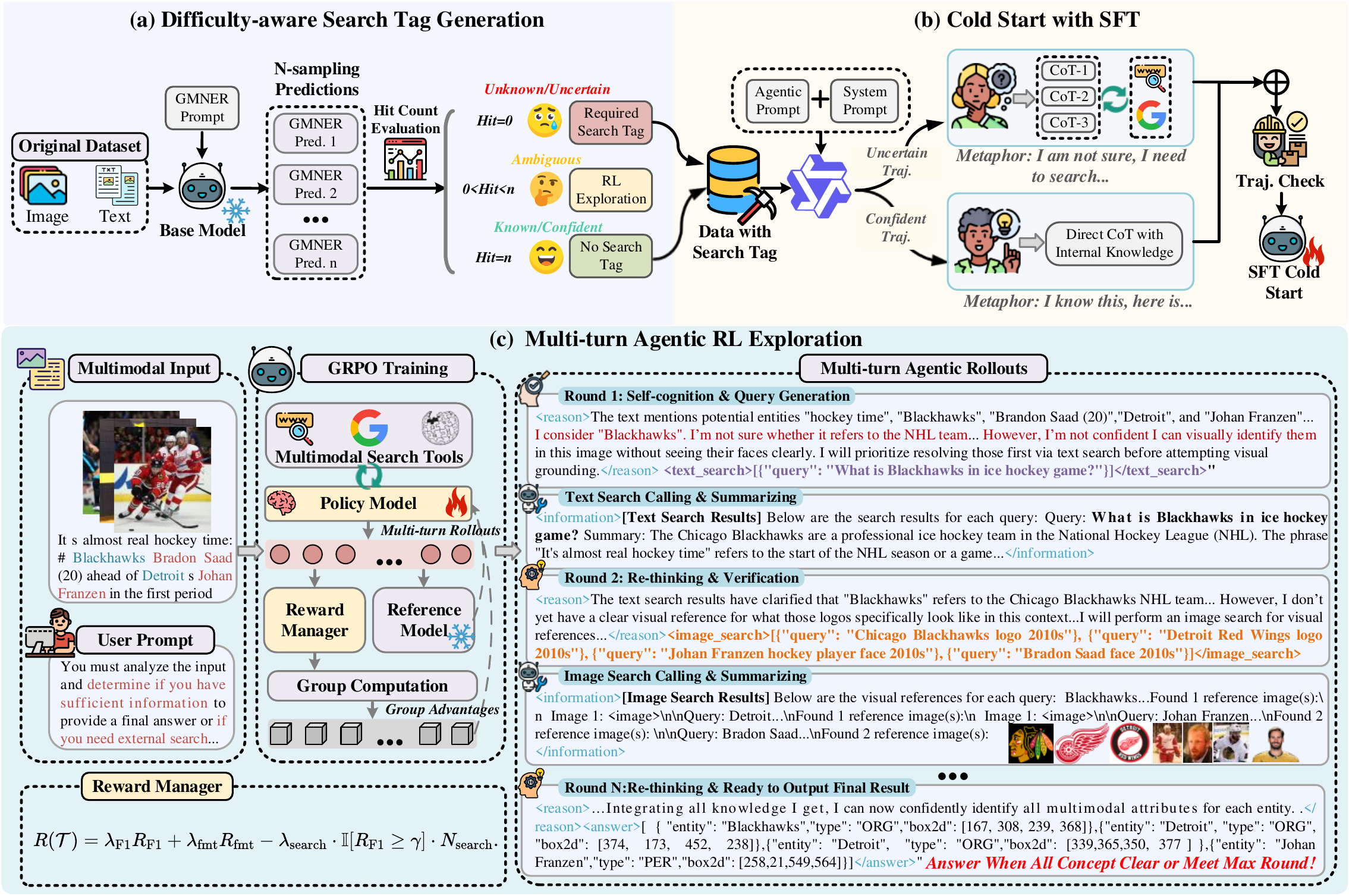}
\caption{Overview of \textit{SAKE}. It presents an agentic framework endowed with explicit self-awareness. SAKE adaptively decides when to invoke external tools for on-demand retrieval and how to generate effective query statements, thereby harmonizing internal reasoning with external knowledge.}
\label{overview}
\end{figure*}

\section{Methodology}
\subsection{Problem Formulation}
Given a textual sequence $T$ and its corresponding image $I$, the goal of Grounded Multimodal Named Entity Recognition (GMNER) is to jointly identify and localize entities by producing a collection of multimodal triplets $Y = \{(s_j, t_j, r_j)\}_{j=1}^{N}$, where $s_j$ denotes the token span of the $j$-th entity in $T$, $t_j$ is its entity type drawn from a predefined category set, and $r_j \in \mathbb{R}^4$ represents the corresponding visual region in $I$. If an entity does not align with any visual region, $r_j$ is set to $\texttt{None}$; otherwise, it specifies the 4D bounding box defined by the top-left and bottom-right coordinates.

\subsection{SAKE}
\subsubsection{Overview.} \textit{SAKE} is an end-to-end agentic multimodal model designed to address GMNER in open-world social media scenarios through self-aware reasoning and adaptive tool invocation. In contrast to prior paradigms of \textit{External Knowledge Exploration} and \textit{Internal Knowledge Exploitation}, \textit{SAKE} is characterized by three core capabilities:
(1) Self-aware reasoning to determine \emph{when} external retrieval is necessary;
(2) Decide \emph{what} information should be retrieved; 
(3) Effectively integrating retrieved external evidence with internal knowledge to produce final GMNER predictions.

\subsubsection{Overall Workflow}
As illustrated in Figure~\ref{overview} (c), given original multimodal inputs together with a user prompt, \textit{SAKE} performs $M$ rounds of reasoning with adaptive tool invocation. In the initial round, the model proposes a set of candidate entities and concepts, conducts Chain-of-Thought reasoning for each candidate, and explicitly expresses its uncertainty. For concepts deemed \emph{uncertain}, \textit{SAKE} generates concrete queries and invokes external search tools.
In subsequent rounds, retrieved evidence is appended to the current reasoning trajectory, enabling the model to reason over the accumulated context. \textit{SAKE} may invoke the search tool multiple times to progressively resolve knowledge gaps. Once all concepts are clear or a predefined maximum number of rounds $M$ is reached, the model produces the final GMNER predictions.

\subsubsection{Action Space.}
At each round, \textit{SAKE} first produces an explicit reasoning step enclosed within
\texttt{<reason>} and \texttt{</reason>}, and then selects one action from the following options:
\begin{itemize}[leftmargin=1.5em]
\item \textbf{Text Search.} Perform batch textual retrieval by generating a
JSON-formatted query enclosed within
\texttt{<text\_search>} and \texttt{</text\_search>}.
\item \textbf{Image Search.} Perform batch image retrieval by generating a
JSON-formatted query enclosed within
\texttt{<image\_search>} and \texttt{</image\_search>}.
\item \textbf{GMNER Prediction.} Produce the final JSON-formatted GMNER prediction
enclosed within \texttt{<answer>} and \texttt{</answer>}.
\end{itemize}

The outputs of the multimodal search tools, enclosed within \texttt{<information>} and \texttt{</information>}, are appended to the current trajectory $\mathcal{T}_t$, resulting in a dynamically evolving reasoning trajectory. Each round must contain exactly one reasoning step and one valid action; trajectories that violate this requirement are considered invalid and discarded.

\begin{figure}[h]
\centering
\includegraphics[width=1.0\linewidth]{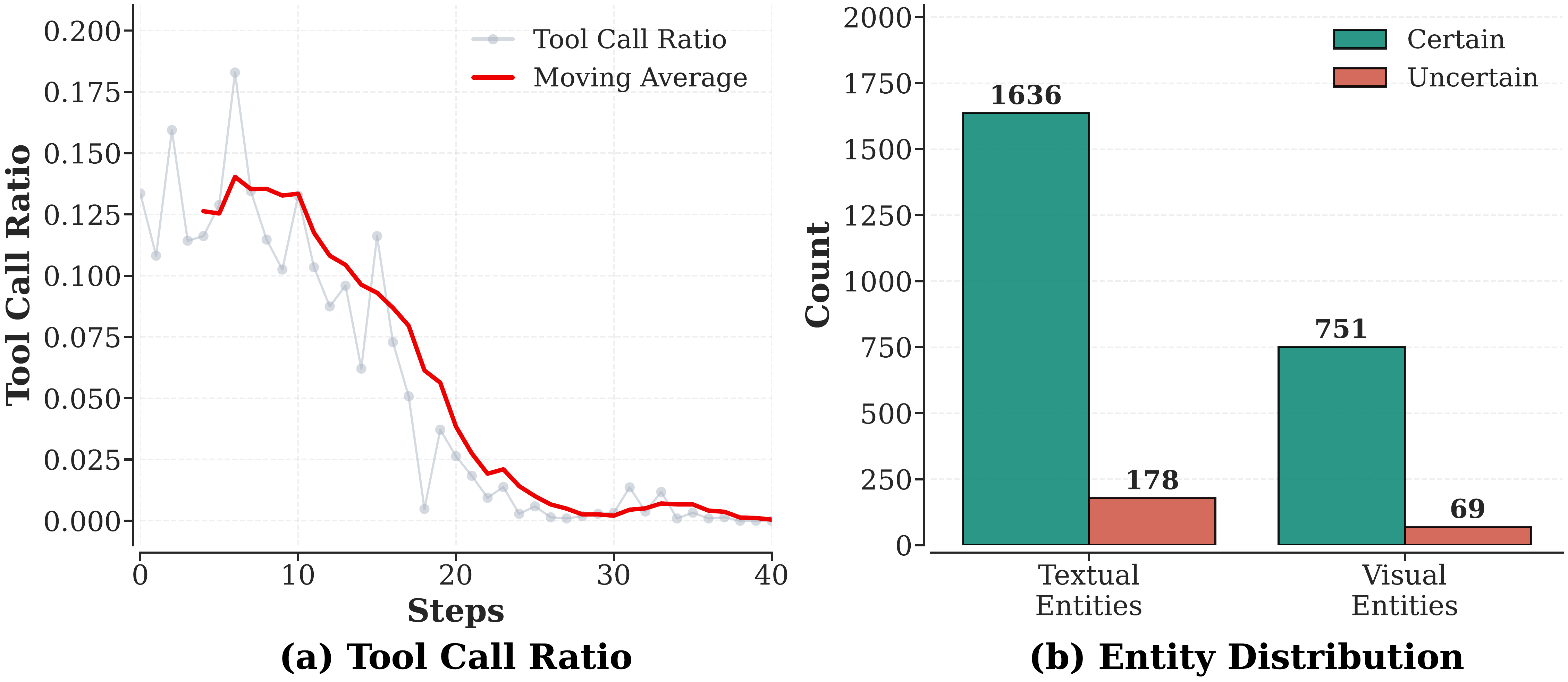}
\caption{(a) Tool invocation ratio without cold-start SFT. (b) Entity prediction uncertainty distribution of the base model.}
\label{tool_call_analysis}
\end{figure}


\subsection{Difficulty-aware Search Tag Generation}
\label{search_tag}
To investigate whether the base model (Qwen2.5-VL-7B-Instruct~\cite{bai2025qwen2}) can identify its knowledge deficiencies and invoke search tools accordingly, we conducted a preliminary experiment using prompt-based self-reflection followed by tool calling accordingly. Following prior works~\cite{jin2025search, wu2025mmsearch, zheng2025deepeyes}, we initially attempted to learn adaptive tool usage relying solely on reinforcement learning. As illustrated in Figure~\ref{tool_call_analysis} (b), when prompted to assess its uncertainty regarding specific visual or textual entities, the model exhibits a distribution heavily skewed toward high certainty. However, this self-assessment correlates poorly with actual performance: as shown in Figure~\ref{sake_behavior_distribution}, the base model exhibits a 63.5\% rate of ``Wrong without Search'' samples, indicating its overconfidence. Consequently, as depicted in Figure~\ref{tool_call_analysis} (a), the model progressively abandons tool invocation during RL training, converging towards a policy that relies exclusively on internal parametric knowledge. This result motivates a cold-start phase that equips the model with explicit awareness of its knowledge gap, enabling targeted queries generation.

Therefore, we propose \textit{Difficulty-aware Search Tag Generation} (see Figure~\ref{overview} (a)).
Specifically, we prompt the base model to perform $N$ forward samplings on the training data. For each ground-truth entity $(s_j, t_j, r_j)$, we compute the model’s hit counts across these predictions $\{(\hat{s}_j^{(k)}, \hat{t}_j^{(k)}, \hat{r}_j^{(k)})\}_{k=1}^{N}$, distinguishing between textual entity hits $Hit_{\text{text}}^{(j)}$ and visual entity hits $Hit_{\text{region}}^{(j)}$:

\begin{equation}
\begin{aligned}
&Hit_{\text{text}}^{(j)} = \sum_{k=1}^{N} \mathbb{I}\big[(\hat{s}_j^{(k)}, \hat{t}_j^{(k)}) = (s_j, t_j)\big], \\
&Hit_{\text{region}}^{(j)} = \sum_{k=1}^{N} \mathbb{I}\big[(\hat{s}_j^{(k)}=s_j)\land (IoU(\hat{r}_j^{(k)} r_j\big)>0.5)\big].
\end{aligned}
\end{equation}

Based on these statistics, we assign a difficulty-aware search tag $\tau_j$ to each ground-truth entity:

\begin{equation}
\tau_j =
\begin{cases}
\texttt{TEXT/IMAGE SEARCH}, 
& Hit_{\text{text}}^{(j)} = 0 \;\text{or}\; Hit_{\text{region}}^{(j)} = 0, \\[4pt]
\texttt{NO\_SEARCH}, 
& Hit_{\text{text}}^{(j)} = N \;\text{and}\; Hit_{\text{region}}^{(j)} = N, \\[4pt]
\texttt{ADAPTIVE}, 
& \text{otherwise}.
\end{cases}
\end{equation}

We define the number of forward samplings $N$ as the \textbf{\textit{Difficulty Level}}. To evaluate the model's grasp of specific entity concepts, we consider two boundary conditions based on the hit count of correct predictions. When the hit count reaches $N$, the model is typically confident about the current entity's attributes and does not require additional retrieval. Conversely, when the hit count is $0$, the model has a vague understanding of the current entity concept and needs to retrieve external knowledge for assistance. These difficulty-aware search tags provide explicit supervision signals during SFT, enabling the model to internalize its knowledge boundaries and to make decisions on when to retrieve external information. Samples with intermediate hit counts are labeled as $\texttt{ADAPTIVE}$ and reserved for RL training, where the model learns to dynamically balance retrieval and internal knowledge.

\subsection{Cold Start with Supervised Fine-tuning}
To equip the base model with fundamental self-awareness and tool-usage capabilities, we leverage the generated search tags to construct chain-of-thought (CoT) trajectories for supervised fine-tuning. To this end, we introduce \textbf{SAKE-SeCoT}, which is used to cold-start the base model.
\subsubsection{SAKE-SeCoT Dataset Construction}
We adopt a three-stage data synthesis pipeline, as illustrated in Figure~\ref{overview} (b). First, we filter samples annotated with \texttt{TEXT SEARCH}, \texttt{IMAGE SEARCH}, and \texttt{NO\_SEARCH} to serve as the initial CoT trajectory samples. Next, we prompt Qwen3-VL-Plus to generate complete solution trajectories by invoking the corresponding tools according to the search tags, explicitly including statements like \textit{``I am not sure, I need to search...''} for entities marked with \texttt{TEXT SEARCH} or \texttt{IMAGE SEARCH}, and \textit{``I know this concept, here is...''} for \texttt{NO\_SEARCH} entities. The results returned by the invoked search tools are fed back iteratively until it produces a final prediction or reaches the maximum number of tool invocations. Finally, we validate the synthesized trajectories with Gemini-2.5-Pro to ensure format correctness, logical consistency, and traceability of retrieval and answers. We filter out trajectories in which the retrieved evidence leads to erroneous predictions or unrelated reasoning paths. This process finally yields $2,764$ high-quality multi-turn CoT trajectories. The statistics of \textbf{SAKE-SeCoT} are presented in Appendix~\ref{appendix:sake_secot}. The prompts for data synthesis are shown in Appendix~\ref{apendix:prompt}.

\subsubsection{Supervised Fine-Tuning}

Given \textbf{SAKE-SeCoT}, we perform supervised fine-tuning (SFT) to initialize the base model with self-aware reasoning and tool-usage behaviors. Each training sample is a multi-turn trajectory $\mathcal{T} = {(o_t, a_t)}_{t=1}^{T}$, where $o_t$ aggregates the multimodal input, retrieved evidence, and preceding context, and $a_t$ denotes the target action, including an explicit reasoning step followed by either a tool invocation or a final GMNER prediction.
The model is trained to autoregressively predict the ground-truth action sequence conditioned on the trajectory prefix. The SFT objective over the dataset $\mathcal{D}_{\text{SeCoT}}$ is:
\begin{equation} \mathcal{L}_{\text{SFT}} = - \mathbb{E}_{\mathcal{T} \sim \mathcal{D}_{\text{SeCoT}}} \sum_{t=1}^{T} \log \pi_{\theta}(a_t \mid o_{\leq t}), \end{equation}
where $\pi_{\theta}$ is the model parameterized by $\theta$.
This stage provides a cold start that equips the model with explicit uncertainty awareness, structured reasoning, and basic tool-invocation skills, forming a foundation for subsequent reinforcement learning.

\subsection{Agentic Reinforcement Learning}
Although SFT endows the model with foundational capabilities for self-awareness and tool calling, the model's rigid imitation of CoT trajectories limits it to suboptimal performance. 
To this end, we further employ agentic RL to enable the model to achieve adaptive reasoning and on-demand tool invocation through interaction with the search environment.

\subsubsection{Rollout with Multi-turn Multimodal Tool Invocation}
\label{sec:rollout}
As illustrated in Figure~\ref{overview} (c), we formalize the rollout process as a multi-turn decision-making where \textit{SAKE} interacts with external environments to progressively resolve uncertainty. Formally, given the multimodal input $x = (T, I)$, the generation process is modeled as a trajectory of interleaved reasoning steps, actions, and environmental observations: $y \sim \pi_\theta(\cdot \mid x; \mathcal{E})$, where $\mathcal{E}$ represents the external search tools. \textit{SAKE} follows an iterative reason-then-action paradigm~\cite{yao2022react}. At each step $t$, the policy $\pi_\theta$ conditions on the accumulated history $H_t$ to generate a response segment $y_t$. This segment contains a chain-of-thought rationale enclosed in \texttt{<reason>} tags, followed by a specific action token. We employ a strictly formatted protocol to trigger tool execution:
\begin{itemize}[leftmargin=1.5em]
    \item \textbf{Tool Invocation:} If the generated sequence $y_t$ contains a search action (i.e., \texttt{<text\_search>} or \texttt{<image\_search>}), the generation is paused. The system parses the JSON query $q$, invokes the corresponding search engine to obtain results $d = \mathcal{E}(q)$, and wraps the evidence within \texttt{<information>} tags. This observation is appended to the history $H_{t+1} \leftarrow H_t \oplus y_t \oplus d$, serving as context for the subsequent round.
    \item \textbf{Termination:} If the model generates the \texttt{<answer>} tag, the process terminates, and the content within is parsed as the final GMNER prediction.
\end{itemize}

This cycle repeats until a valid answer is produced or the maximum action budget $M$ is exhausted. The detailed workflow is outlined in Algorithm~\ref{alg:rollout}. The complete prompts are provided in Appendix~\ref{apendix:prompt}.

\subsubsection{Group Relative Policy Optimization}
To ensure training stability and computational efficiency, we employ Group Relative Policy Optimization (GRPO)~\cite{shao2024deepseekmath}, which eliminates the need for an additional value function approximation (critic model). Specifically, for each multimodal input $x$, \textit{SAKE} samples a group of trajectories $\{y_1, \dots, y_G\}$ from the current policy $\pi_{\theta}$. Instead of relying on a learned value baseline, GRPO estimates the advantage of each trajectory by normalizing its reward against the group statistics. This group-relative baseline effectively reduces gradient variance and encourages the model to generate reasoning paths that outperform the average of the sampled group. Furthermore, since our trajectories involve external tool interactions, we apply specific token masking to exclude deterministic search results from the gradient computation. The detailed objective function and derivation are provided in Appendix~\ref{app:grpo}.

\subsubsection{Reward Modeling}
\label{sec:reward_modeling}
To align the model with accurate, structured, and efficient reasoning, we define the reward function $R(\mathcal{T})$ for a trajectory $\mathcal{T}$ as:
\begin{equation}
    R(\mathcal{T}) = \lambda_{\text{F1}} R_{\text{F1}} + \lambda_{\text{fmt}} R_{\text{fmt}} - \lambda_{\text{search}} \cdot \mathbb{I}[R_{\text{F1}} \geq \gamma] \cdot N_{\text{search}}.
\end{equation}
Here, $R_{\text{F1}}$ denotes the F1 score of the predicted multimodal entities against ground truth, ensuring task accuracy. $R_{\text{fmt}}$ is a binary reward indicating compliance with the required action tag (e.g., \texttt{<reason>}, \texttt{<text\_search>}, etc.), which is essential for parsing. The final term introduces an efficiency penalty based on the average search count per entity ($N_{\text{search}}$). Crucially, this penalty is gated by an indicator function $\mathbb{I}[R_{\text{F1}} \geq \gamma]$; it activates only when the prediction quality meets a threshold $\gamma$. This mechanism encourages the model to explore tools freely during early learning stages but penalizes redundant retrieval once high accuracy is achieved, effectively balancing the trade-off between performance and retrieval cost.

\begin{table*}[h]
\caption{Comparisons of competitive methods on two GMNER datasets. 
Bold and underlined results indicate the best and second-best performance, respectively. 
E2E denotes end-to-end GMNER model without intermediate task-specific modules, and Avg. SR denotes the average search ratio. 
$^\ast$ indicates methods re-implemented under settings consistent with SAKE.
}
\centering
\small
\begin{adjustbox}{width=2.0\columnwidth}
\begin{tabular}{c|l|ccc|ccc|ccc}
\toprule
\multirow{2}{*}{\textbf{Category}} 
& \multirow{2}{*}{\textbf{Methods}} 
& \multirow{2}{*}{\textbf{E2E}} 
& \multirow{2}{*}{\textbf{Using LLM}} 
& \multirow{2}{*}{\textbf{Avg. SR (\%)}} 
& \multicolumn{3}{c|}{\textbf{Twitter-GMNER}} 
& \multicolumn{3}{c}{\textbf{Twitter-FMNERG}} \\
& & & & 
& GMNER & MNER & EEG 
& GMNER & MNER & EEG \\
\midrule

\multirow{3}{*}{\textit{Few-shot MLLM}} 
& GPT4o \cite{hurst2024gpt}  
& \Checkmark & GPT4o & 0.0 
& 41.29 & 65.07 & 44.95 
& 32.37 & 52.26 & 41.60 \\

& Gemini-2.5 Pro \cite{comanici2025gemini} 
& \Checkmark & Gemini-2.5 Pro & 0.0 
& 43.31 & 64.57 & 47.63 
& 34.02 & 51.14 & 45.89 \\

& Qwen3-VL-Plus \cite{bai2025qwen2} 
& \Checkmark & Qwen3-VL-Plus & 0.0 
& 45.17 & 66.80 & 48.72 
& 31.43 & 48.57 & 47.04 \\

\midrule

\multirow{8}{*}{\makecell{\textit{Supervised}\\\textit{Fine-tuning}}}
& UMT-OD-EVG \cite{umt} 
& \XSolidBrush & - & 0.0 
& 50.29 & 78.58 & 54.78 
& 41.32 & 61.63 & 54.43 \\

& UMGF-OD-EVG \cite{umgf} 
& \XSolidBrush & - & 0.0 
& 51.67 & 78.83 & 55.74 
& 41.92 & 61.79 & 54.75 \\

& ITA-OD-EVG \cite{wang2022ita} 
& \XSolidBrush & - & 0.0 
& 51.56 & 79.37 & 55.69 
& 42.78 & 63.21 & 57.26 \\

& MMT5 / BART-OD-EVG \cite{gmner} 
& \XSolidBrush & - & 0.0 
& 52.45 & 80.39 & 55.66 
& 45.21 & 66.61 & 58.18 \\

& H-Index \cite{gmner} 
& \Checkmark & - & 0.0 
& 56.41 & 79.73 & 61.18 
& 46.55 & 64.84 & 60.46 \\

& TIGER \cite{fg-gmner} 
& \Checkmark & - & 0.0 
& 57.48 & - & - 
& 47.20 & 64.91 & 61.96 \\

& MQSPN \cite{mqspn} 
& \Checkmark & - & 0.0 
& 58.76 & 80.43 & 62.40 
& 47.86 & 66.83 & 61.95 \\

& Qwen2.5-VL-7B (SFT)$^\ast$ \cite{bai2025qwen2} 
& \Checkmark & Qwen2.5-VL-7B & 0.0 
& 68.49 & 82.31 & 71.66
& 58.20 & 68.93 & 72.01 \\

\midrule

\multirow{3}{*}{\makecell{\textit{External Knowledge}\\\textit{Exploration}}} 

& SCANNER \cite{ok2024scanner} 
& \XSolidBrush & BLIP-2 & 100.0 
& 68.52 & - & - 
& - & - & - \\

& VKEL \cite{yuan2026visual} 
& \XSolidBrush & LLaVA-1.5-7B & 100.0 
& 66.87 & 83.75 & 70.43 
& 57.44 & \underline{72.59} & 68.46 \\

& Qwen2.5-VL-7B (MoRE)$^\ast$ \cite{wang2022named} 
& \XSolidBrush & Qwen2.5-VL-7B & 100.0
& 69.01 & 82.67 & 72.25
& 58.37 & 69.14 & 72.38 \\

\midrule

\multirow{5}{*}{\makecell{\textit{Internal Knowledge}\\\textit{Exploitation}}}

& GEM \cite{wang2024granular} 
& \XSolidBrush & ChatGPT & 0.0
& 61.54 & 84.81 & 64.49 
& 52.48 & 70.80 & 65.52 \\

& UnCo \cite{tang2025-unco} 
& \XSolidBrush & Gemini-2.5 Pro & 0.0 
& 64.58 & 81.71 & 69.62
& 53.56 & 67.70 & 68.25 \\

& RiVEG \cite{li2024llms} 
& \XSolidBrush & ChatGPT & 0.0 
& 67.06 & 85.94 & 70.01
& - & - & - \\

& PGIM-ISR-OFA \cite{yu2025isr} 
& \XSolidBrush & Qwen2-VL-7B & 0.0
& \underline{72.35} & - & -
& - & - & - \\

& Qwen2.5-VL-7B (SFT+GRPO)$^\ast$ \cite{bai2025qwen2}
& \Checkmark & Qwen2.5-VL-7B & 0.0 
& 69.74 & 82.58 & 73.08
& 59.91 & 69.57 & 73.80\\

\midrule

& MAKAR \cite{lin2025makar} 
& \XSolidBrush & Qwen2.5-VL-7B & 100.0 
& {71.88} & \underline{86.38} & \underline{74.64}
& \underline{60.54} & {71.24} & \underline{75.66} \\

\rowcolor{blue!10} 
\cellcolor{white}  
\multirow{-2}{*}{\textit{LLM Agent}} 
& \textbf{SAKE (Ours)}
& \Checkmark & Qwen2.5-VL-7B & 68.8 
& \textbf{75.63 }& \textbf{86.95} & \textbf{78.76 }
& \textbf{63.45} & \textbf{73.37} & \textbf{77.12} \\
\bottomrule
\end{tabular}
\end{adjustbox}
\label{tab:main}
\end{table*}

\begin{table*}[h]
\caption{Ablation study results on Seen and Unseen samples across two datasets. Unseen refers to samples containing entity attributes (entity text mention, type, or region) that did not appear in the training set; otherwise, denoted as Seen.}
\centering
\small
\begin{adjustbox}{width=2.0\columnwidth}
\begin{tabular}{l|cccccc|cccccc}
\toprule
\multirow{3}{*}{\textbf{Settings}}  & \multicolumn{6}{c|}{\textbf{Twitter-GMNER}} & \multicolumn{6}{c}{\textbf{Twitter-FMNERG}} \\
\cline{2-13} 

 & \multicolumn{2}{c|}{\textbf{Seen}} & \multicolumn{2}{c|}{\textbf{Unseen}} & \multicolumn{2}{c|}{\textbf{All}} & \multicolumn{2}{c|}{\textbf{Seen}} & \multicolumn{2}{c|}{\textbf{Unseen}} & \multicolumn{2}{c}{\textbf{All}} \\ 
 
 & {SR (\%)} & \multicolumn{1}{c|}{{GMNER}} & {SR (\%)} & \multicolumn{1}{c|}{{GMNER}} & {SR (\%)} & \multicolumn{1}{c|}{{GMNER}} & {SR (\%)} & \multicolumn{1}{c|}{{GMNER}} & {SR (\%)} & \multicolumn{1}{c|}{{GMNER}} & {SR (\%)} & \multicolumn{1}{c}{{GMNER}} \\
\midrule
(1) w/o Cold Start & 0.0 & 77.15 & 0.0 & 52.62 & 0.0 & 64.26 & 0.0 & 67.07 & 0.0 & 43.29 & 0.0 & 47.26 \\
(2) w/o Search Tag & 0.0 & 80.52 & 0.0 & 58.48 & 0.0 & 69.90 & 0.8 & 73.12 & 1.0 & 47.70 & 0.9 & 59.52\\
(3) w/o GRPO  & 85.6 & 76.88 & 88.9  & 49.39 & 87.4 & 61.87 & 83.0 & 69.65 & 87.1 & 44.48 & 85.8 & 49.99\\
(4) w/o Search Penalty & 79.8 & 84.57 & 85.0 & 66.30 & 82.6 & 75.92 & 75.2 & 76.24 & 81.9 & 52.94 & 79.5 & 63.82\\
(5) w/o Image Search & 58.2 & 82.16 & 68.8 & 62.25 & 63.9 & 72.18 & 60.7 & 73.38 & 68.3 & 50.13 & 65.6 & 61.04\\
(6) w/o Text Search & 6.9 & 81.03 & 19.5 & 59.78 & 13.7 & 70.46 & 11.3 & 71.67 & 25.6  & 45.85 & 20.5 & 57.81\\
\midrule
\rowcolor{blue!10} 
SAKE & 59.4 & 84.34 & 72.2 & 66.07 & 66.3 & 75.63 & 64.4 & 75.96 & 75.1 & 52.57 & 71.3 & 63.45  \\
\bottomrule
\end{tabular}
\end{adjustbox}
\label{tab:ablation}
\end{table*}

\section{Experiments}
\subsection{Experiment Settings}
\textbf{GMNER Benchmarks \& Evaluation} We evaluate our approach on two benchmarks, Twitter-GMNER~\cite{gmner} and Twitter-FMNERG~\cite{fg-gmner}. The details of datasets are provided in Appendix~\ref{apendix:datasets}. We follow the standard evaluation protocol for GMNER as described in~\cite{gmner}, assessing performance on both Multimodal Named Entity Recognition (MNER) and Entity Extraction and Grounding (EEG).  Overall performance is reported using F1 scores. Detailed metric definitions are provided in Appendix~\ref{apendix:metrics}.


\textbf{Implementation Details.} \textit{SAKE-7B} is built upon Qwen2.5-VL-7B-Instruct and trained using a two-stage pipeline. All experiments are conducted on 8 NVIDIA H100 GPUs with 80GB memory. The SFT stage is implemented with the \textit{ms-swift}~\cite{zhao2025swift} framework, while the RL stage is conducted using \textit{veRL}~\cite{sheng2025hybridflow}. 
During SFT, the model is trained on SAKE-SeCoT to acquire instruction-following capability, awareness of its own knowledge gaps, and the ability to invoke external tools. In the subsequent RL stage, a reward model guides the model to balance internal knowledge utilization and external knowledge exploration. 
Detailed training hyperparameters and search tool are provided in Appendix~\ref{apendix:implementation}.



\textbf{Baselines.} We group existing baseline methods into five categories.
(i) \textbf{Few-shot MLLM} methods, where MLLMs perform direct GMNER prediction via 3-shot in-context learning~\cite{min2022rethinking}.
(ii) \textbf{Supervised fine-tuning} methods, which learn to directly predict GMNER outputs from annotated training data.
(iii) \textbf{External knowledge exploration} methods, where models augment their inputs with retrieved evidence from search engines and leverage the additional context for supervised fine-tuning or direct prompting.
(iv) \textbf{Internal knowledge exploitation} methods, which rely on the internal knowledge of MLLMs to refine GMNER results. (v) \textbf{LLM Agent} methods, which tackle unseen and long-tail entity recognition in GMNER using multiple LLM agents and external tools. 
Detailed descriptions of the baselines and prompt templates are provided in Appendix~\ref{appendix:baseline} and Appendix~\ref{apendix:prompt}, respectively.

\subsection{Main Results}

Table \ref{tab:main} presents the performance comparison between SAKE and baseline methods on the GMNER task and its two subtasks (MNER and EEG) across two datasets. Based on these results, we derive the following key observations: \textbf{(1) Generalization via Reasoning and Exploration.} 
Compared to supervised fine-tuning (SFT) methods, LLM-based approaches leveraging internal reasoning and external knowledge exploration significantly enhance generalization in social media GMNER. \textbf{(2) Superiority over External Knowledge Exploration.} 
SAKE outperforms all methods relying on external knowledge exploration. This aligns with our expectation that indiscriminate retrieval introduces excessive noise which degrades the performance on known entity concepts. In contrast, SAKE effectively boosts performance by maintaining self-awareness of its knowledge deficiencies and invoking search only when necessary. \textbf{(3) Superiority over Internal Knowledge Exploitation.} 
SAKE surpasses all internal knowledge exploitation methods based on RL training and self-refinement. These results expose the limitations of relying solely on LLM parametric knowledge in open-world social media scenarios. Strategically utilizing external search to address knowledge gaps proves crucial for performance gains. \textbf{(4) New State-of-the-Art Performance.} SAKE consistently achieves state-of-the-art (SOTA) results. Compared to the previous SOTA, MAKAR, SAKE achieves improvements of 3.75\% and 2.91\% on the GMNER and FMNERG datasets, respectively. Notably, unlike MAKAR, which employs a multi-agent system with full-scale retrieval, SAKE operates as an efficient end-to-end agent. It surpasses MAKAR with fewer intermediate steps and a significantly reduced average search rate of 68.8\%, further demonstrating its superiority and efficiency.

\subsection{Ablation Study}
\textbf{Ablation Setting.} To validate the effectiveness of SAKE's module design at different stages, we conducted a comprehensive ablation study on two datasets. This includes an analysis of SAKE's performance and the variations in its search tool invocation behavior across \textit{Seen}, \textit{Unseen}, and \textit{All} samples. Specifically, during the SFT phase: \textbf{(1) w/o Cold Start}: We removed the cold start phase and directly trained the model using GRPO. \textbf{(2) w/o Search Tag:} We eliminated the search tag and directly used the CoT trajectories generated by the teacher model (Qwen3-VL-Plus) for supervised fine-tuning of SAKE. During the RL phase: \textbf{(3) w/o GRPO:} We removed the GRPO training and directly used SFT for training SAKE. \textbf{(4) w/o Search Penalty:} We removed the search penalty and calculated the advantage using only the F1 reward and format reward. For multimodal search tools, \textbf{(5) w/o Image Search:} We excluded the image search tool and only used the text search tool for RL training. \textbf{(6) w/o Text Search:} We removed the text search tool and solely relied on the image search tool for RL training. The experimental results are shown in Table~\ref{tab:ablation}.

\textbf{SFT Training Phase.} In setting (1), we observe a significant degradation in GMNER performance across all samples upon removal of the cold-start phase, accompanied by a multimodal search rate of zero. This indicates that relying solely on RL exploration is insufficient to elicit the search tool invocation capabilities of the base model (Qwen2.5-VL-7B). Cold start is crucial for enhancing model performance and tool invocation capabilities. In setting (2), omitting search tags while distilling knowledge from the teacher model's CoT trajectories still results in low tool-use rate. This is primarily attributed to the teacher model's extensive parametric knowledge, which encourages reliance on internal reasoning rather than external tool invocation. Conversely, the Difficulty-aware Search Tag serves as an explicit indicator of knowledge gaps, enabling the base model to recognize its limitations and trigger necessary tool invocations.

\textbf{RL Training Phase.} In setting (3), the ablation results indicate that removing GRPO leads to an obvious decline in performance; specifically, GMNER and FMNERG on Unseen samples drop by 16.68\% and 8.09\%, respectively. Paradoxically, the search ratio increases. This suggests that RL is thus essential for teaching the model precisely \textit{when} to retrieve and \textit{what} content to query. Furthermore, in setting (4), the incorporation of a search penalty effectively balances retrieval efficiency with performance, mitigating tool misuse and redundancy. Additionally, experimental results reveal that RL significantly increases the search ratio for Unseen samples relative to Seen samples. This further corroborates that RL empowers the model to optimize the decision-making between exploiting internal parametric knowledge and exploring external information.

\textbf{Multimodal Search Tools.} We further investigate the impact of distinct search modalities on SAKE's performance. Excluding the image search tool causes GMNER and FMNERG scores to decline by 3.45\% and 2.41\%, respectively. In comparison, removing the text search tool leads to more pronounced drops of 5.17\% and 5.64\%. These findings confirm that while both search modalities confer positive gains, text retrieval yields a superior performance boost. We attribute this disparity to the fact that current multimodal large Language models remain text-dominant within interleaved multimodal Chain-of-Thought reasoning. Effectively leveraging visual information requires ``thinking with images'' capability which remains a significant challenge for existing architectures.

\begin{figure}[]
\centering
\includegraphics[width=1.0\linewidth]{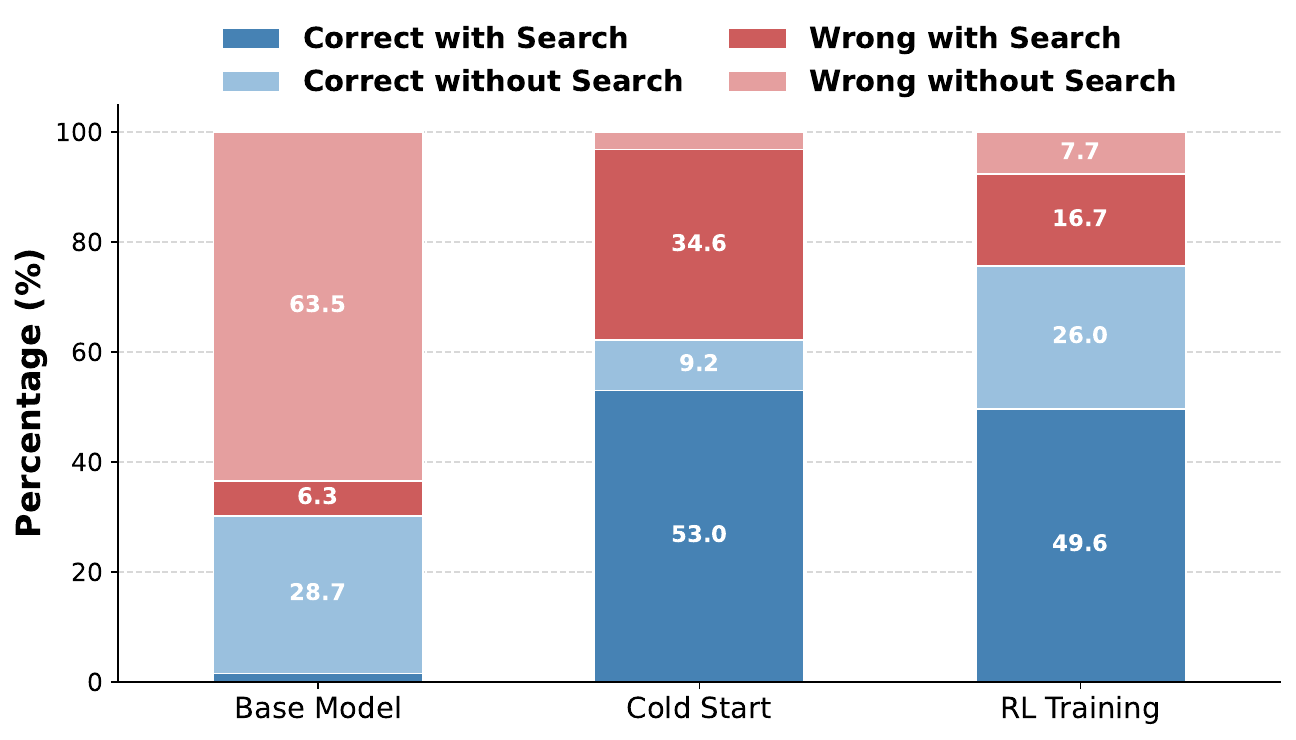}
\caption{Evolution of behavioral distribution across two training stages (SFT+RL) and base model (Qwen2.5-VL-7B).}
\label{sake_behavior_distribution}
\end{figure}

\begin{figure}[]
\centering
\includegraphics[width=1.0\linewidth]{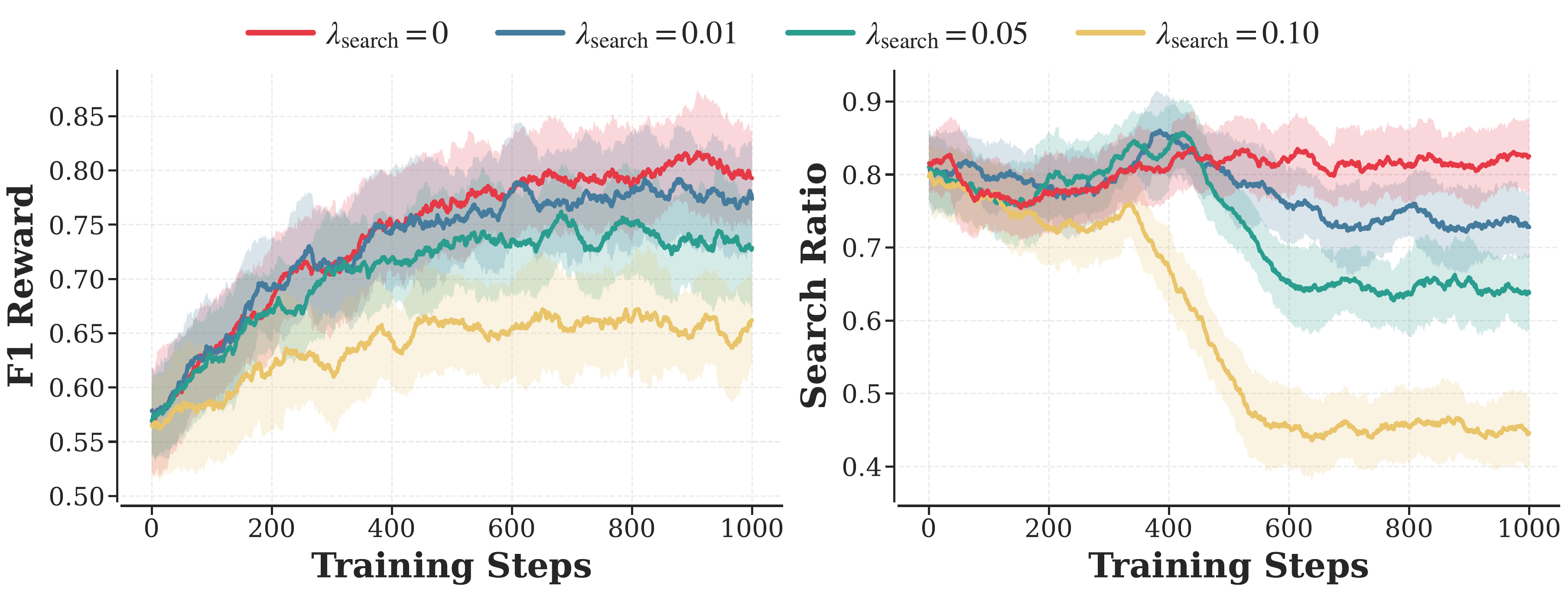}
\caption{RL Training Dynamics on F1 Reward and Search Ratio under Different Search Penalty $\lambda_{\text{search}}$.}
\label{RL_dynamic_comparison}
\end{figure}

\subsection{Discussion and Analysis}
\textbf{Evolution of SAKE Capabilities Across Two-Stage Training.} Figure~\ref{sake_behavior_distribution} illustrates SAKE's capabilities evolution. Initially, the \textbf{Base model} exhibits cognitive limitations, with a high error rate (63.5\% ``Wrong without Search'') due to its inability to recognize knowledge boundaries, resulting in minimal search tool invocations. The \textbf{Cold Start (SFT)} stage introduces search awareness, improving the ``Correct with Search'' rate to 53.0\%. However, this also leads to over-reliance on external tools, causing the internal confidence (``Correct without Search'') to plummet to 9.2\%, while noise-induced errors increase. This behavior is primarily due to the model's tendency to mimic the search action in the SAKE-SeCoT dataset, rather than adaptively  assessing the necessity of retrieval. The final \textbf{RL stage} rectifies this by penalizing redundant retrieval. This optimization restores reliance on internal memory (rebounding ``Correct without Search'' to 26.0\%) and minimizes search noise (dropping ``Wrong with Search'' to 16.7\%), allowing SAKE to dynamically arbitrate between internal knowledge exploitation and external exploration.

\textbf{Impact of the Search Penalty $\lambda_{\text{search}}$ on RL Training Dynamics.} To elucidate the influence of the search penalty $\lambda_{\text{search}}$ on RL training, Figure~\ref{RL_dynamic_comparison} illustrates the training dynamics of the F1 Reward and Search Ratio across varying penalty magnitudes. We observe that in the absence of a search penalty ($\lambda_{\text{search}}=0$), the F1 reward converges to approximately 0.8, with the search ratio similarly oscillating around 0.8. While yielding the highest F1 score, this setting incurs substantial search costs. As $\lambda_{\text{search}}$ is increased to 0.01, 0.05, and 0.10, the F1 score exhibits a progressive decline. This highlights an inherent trade-off: a larger penalty corresponds to reduced F1 performance but significantly lower search consumption. Notably, with $\lambda_{\text{search}}$ at the magnitude of $10^{-2}$, the model maintains a comparable F1 score while achieving a significantly lower and more stable search ratio. This suggests that the model has learned to invoke search tools only when necessary, thereby achieving on-demand search behavior.

\begin{figure}[t]
\centering
\includegraphics[width=1.0\linewidth]{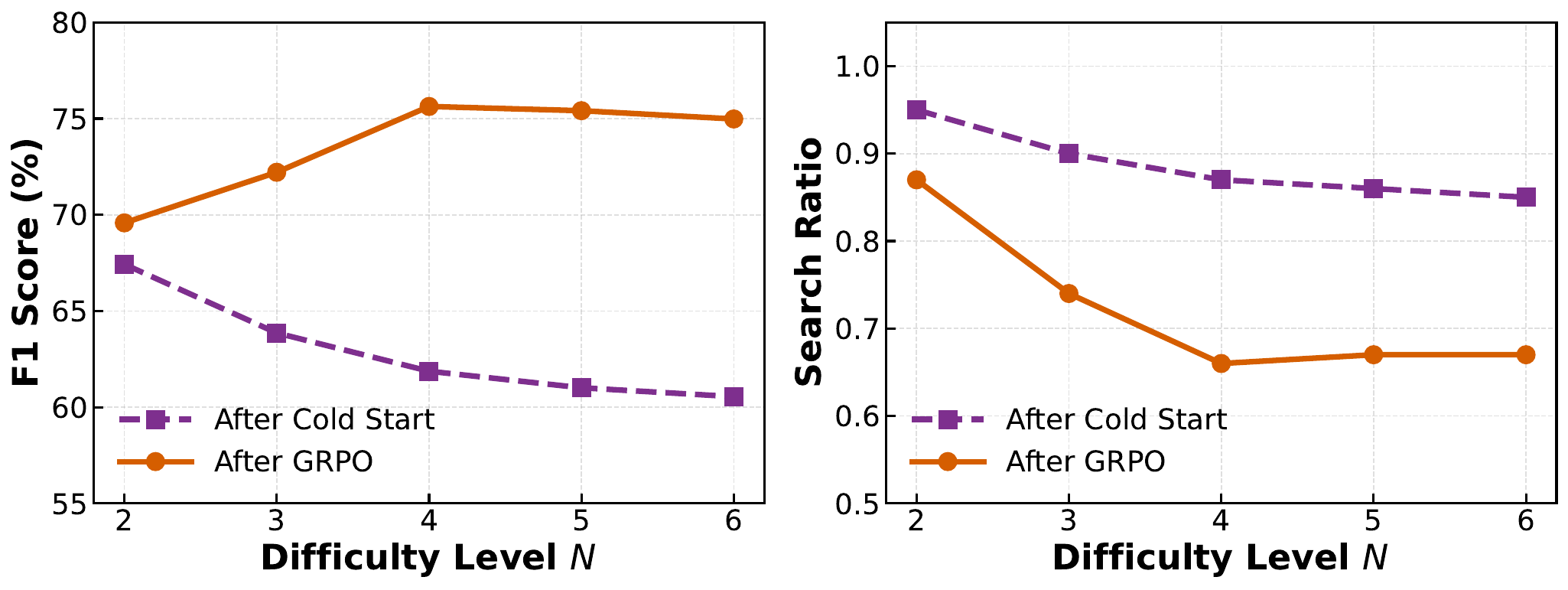}
\caption{Performance and Search Ratio under Different Difficulty Level $N$ of Search Tags.}
\label{search_tag_difficulty}
\end{figure}

\textbf{Impact of the Difficulty Level $N$ on Search Tag Generation.} To investigate the impact of the Difficulty Level $N$ on performance and search behaviors, we generated search tags across varying levels of $N$ for both the cold start and RL training phases, as shown in Figure~\ref{search_tag_difficulty}. We observe that as $N$ increases, the search ratio in both phases exhibits a consistent decline, eventually stabilizing around $N=4$. This trend is primarily attributed to the fact that a larger $N$ increases the probability of the model successfully hitting ground truth entities during multiple forward sampling. Consequently, a larger proportion of training samples are labeled as $\texttt{ADAPTIVE}$ for RL, while the frequency of \texttt{SEARCH} tags diminishes. Furthermore, while a higher $N$ leads to a reduction in F1 scores during the cold start phase, it significantly boosts performance in the subsequent RL stage. To achieve a balanced data distribution between the cold start and RL phases, we adopt Difficulty Level $N=4$ as the optimal criterion for Search Tag Generation in this work. \textbf{Please refer to Appendix~\ref{appendx:experiment} for more experimental results and analysis.}

\section*{Conclusion}
In this paper, we presented \textit{SAKE}, an end-to-end agentic framework designed to address the challenges of open-world Grounded Multimodal Named Entity Recognition (GMNER). Unlike existing approaches that rely rigidly on either internal parametric memory or heuristic external retrieval, SAKE harmonizes these two paradigms through self-aware reasoning and adaptive retrieval. By introducing Difficulty-aware Search Tag Generation and a two-stage training paradigm combining supervised fine-tuning with reinforcement learning, we effectively equip MLLMs with the ability to thinking their knowledge deficiencies and perform adaptive, on-demand information seeking. Extensive experiments on two widely used benchmarks demonstrate that SAKE achieves state-of-the-art performance while significantly reducing search frequency. 




  \bibliographystyle{ACM-Reference-Format}
  \bibliography{sample-base}

\appendix

\section{Statistics of SAKE-SeCoT}
\label{appendix:sake_secot}
\begin{figure*}[]
\centering
\includegraphics[width=1.0\linewidth]{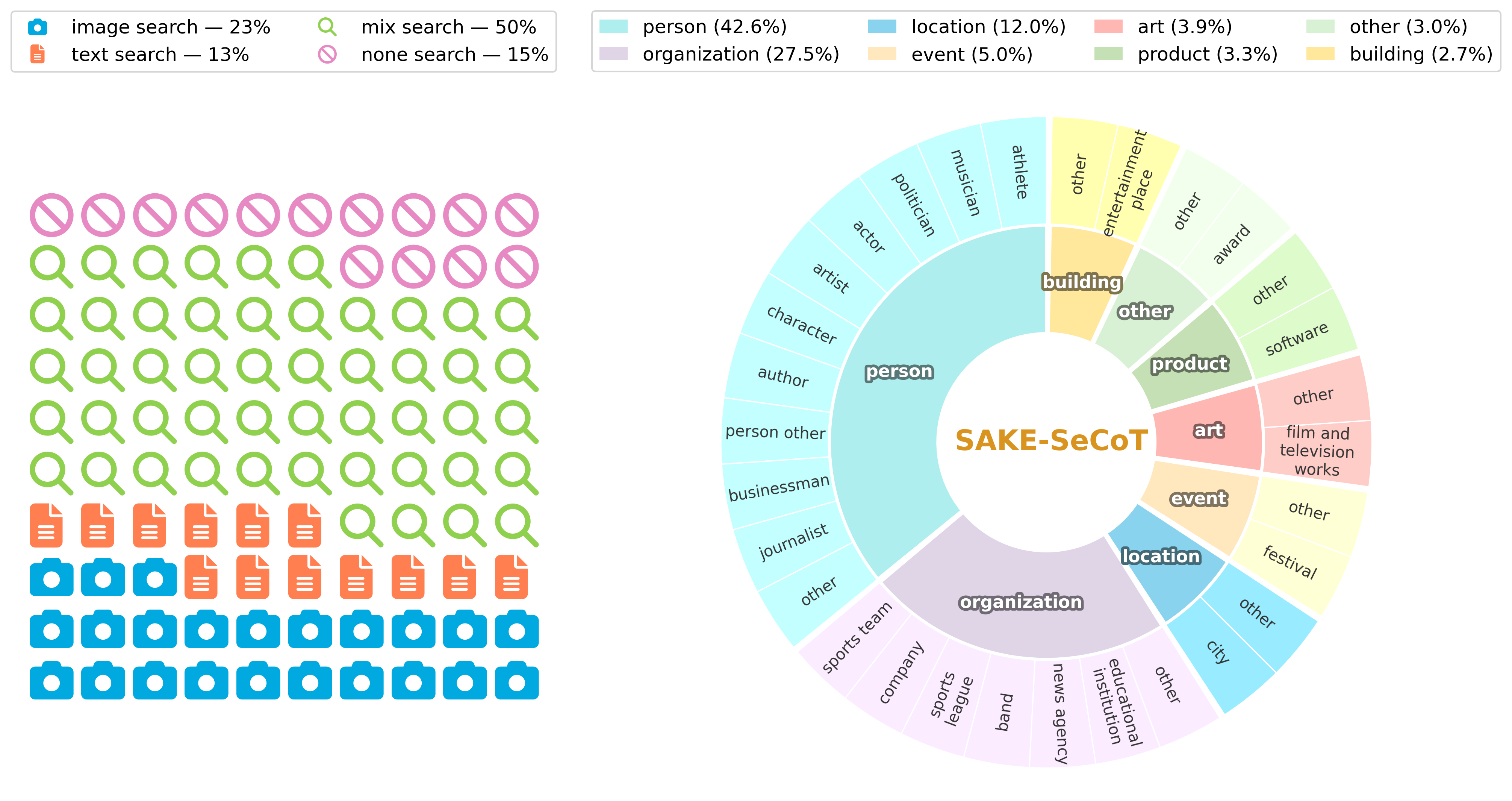}
\caption{The Statistics of SAKE-SeCoT. The left plot shows the distribution of search behaviors in the dataset, while the right plot illustrates the distribution of entities across coarse-grained and fine-grained types.}
\label{SAKE_SeCoT}
\end{figure*}

The SAKE-SeCoT dataset (see Figure~\ref{SAKE_SeCoT}) is designed to enable the base model to recognize its own knowledge gaps and develop the ability to generate appropriate queries for tool invocation. It consists of 2,764 high-quality multi-turn Chain-of-Thought (CoT) trajectories, each containing between 0 and 18 entity mentions. The left plot shows the distribution of search tags assigned to entities, revealing that mixed (text and image) search is used in approximately half of the cases. The right plot categorizes entities into eight coarse-grained types and 51 fine-grained categories, covering a range of entities including people, organizations, locations, events, products, and more. This diversity supports research on selective retrieval and enhances the robustness of open-world entity recognition.

\section{SAKE Rollout with Multi-turn Tool Invocation}
Algorithm~\ref{alg:rollout} details the multi-round tool invocation process of SAKE. Unlike previous R1-based reasoning models, SAKE adaptively analyzes knowledge gaps in multimodal inputs, generates corresponding queries, and dynamically invokes external search tools. Upon receiving the results, it evaluates whether to output the final result or proceed with another round of interaction. The entire process can be viewed as a multi-turn dialogue with the external search environment, terminating either when the predefined turn limit is reached or the \texttt{<answer>} token is output.

\begin{algorithm}[h]
\caption{\textit{SAKE}'s Rollout with Multi-turn Tool Invocation}
\label{alg:rollout}
\begin{algorithmic}[1]
\REQUIRE Multimodal input $x$, Policy $\pi_\theta$, Search Tools $\mathcal{E}$, Max action budget $M$.
\ENSURE Final GMNER prediction $y$.
\STATE Initialize history $H \leftarrow x$
\STATE Initialize action count $m \leftarrow 0$
\WHILE{$m < M$}
    \STATE Initialize current step sequence $y_t \leftarrow \varnothing$
    \STATE \COMMENT{Autoregressive Generation}
    \WHILE{True}
        \STATE Sample token $w_i \sim \pi_\theta(\cdot \mid H \oplus y_t)$
        \STATE $y_t \leftarrow y_t \oplus w_i$
        \IF{$w_i$ completes a valid closing tag (\texttt{</text\_search>}, \texttt{</image\_search>}, or \texttt{</answer>})}
            \STATE \textbf{break}
        \ENDIF
    \ENDWHILE
    
    \STATE \COMMENT{Action Execution and Environment Interaction}
    \IF{\texttt{<text\_search>} or \texttt{<image\_search>} detected in $y_t$}
        \STATE Extract query $q \leftarrow \text{Parse}(y_t)$
        \STATE Execute search $d \leftarrow \mathcal{E}(q)$
        \STATE Format observation $obs \leftarrow \texttt{<information>} \oplus d \oplus \texttt{</information>}$
        \STATE Update history $H \leftarrow H \oplus y_t \oplus obs$
        \STATE $m \leftarrow m + 1$
    \ELSIF{\texttt{<answer>} detected in $y_t$}
        \STATE Extract final prediction $y \leftarrow \text{ParseAnswer}(y_t)$
        \RETURN $y$
    \ELSE
        \STATE \COMMENT{Handle invalid formats or hallucinated actions}
        \STATE $H \leftarrow H \oplus y_t \oplus \text{``Invalid Action. Please retry.''}$
    \ENDIF
\ENDWHILE
\RETURN Empty or partial prediction (Budget Exhausted)
\end{algorithmic}
\end{algorithm}

\section{Details of Group Relative Policy Optimization in SAKE}
\label{app:grpo}

In this section, we provide the detailed formulation of the GRPO objective used in SAKE.

\paragraph{Objective Function.}
For each input $x$, we sample a group of $G$ outputs $\{y_1, y_2, \dots, y_G\}$ from the old policy $\pi_{\theta_{\text{old}}}$. The policy model is optimized by maximizing the following objective:

\begin{equation}
\resizebox{\columnwidth}{!}{$
\begin{aligned}
\mathcal{J}_{\text{GRPO}}(\theta) = \mathbb{E}_{x \sim \mathcal{D}, \{y_i\}_{i=1}^G \sim \pi_{\theta_{\text{old}}}(\cdot|x)} \bigg[ \frac{1}{G} \sum_{i=1}^G \bigg( 
& \frac{1}{|y_i|} \sum_{t=1}^{|y_i|} 
\min \left( 
\rho_{i,t} \hat{A}_{i}, 
\text{clip}(\rho_{i,t}, 1-\epsilon, 1+\epsilon) \hat{A}_{i} 
\right) \\
& - \beta \mathbb{D}_{\text{KL}} \left( 
\pi_\theta(\cdot|x, y_{i,<t}) \| 
\pi_{\text{ref}}(\cdot|x, y_{i,<t}) 
\right) 
\bigg) \bigg]
\end{aligned}
$}
\end{equation}

where $\rho_{i,t} = \frac{\pi_\theta(y_{i,t} | x, y_{i,<t})}{\pi_{\theta_{\text{old}}}(y_{i,t} | x, y_{i,<t})}$ denotes the importance sampling ratio, and $\epsilon$ is the clipping hyperparameter. The term $\beta$ controls the strength of the KL divergence penalty, ensuring the trained policy does not deviate excessively from the reference SFT model $\pi_{\text{ref}}$.

\paragraph{Advantage Estimation.}
The advantage $\hat{A}_{i}$ for the $i$-th trajectory is computed based on the relative rewards within the group. Let $r_i$ be the reward for trajectory $y_i$ (computed as described in Sec.~\ref{sec:reward_modeling}). The advantage is standardized as:
\begin{equation}
\hat{A}_i = \frac{r_i - \text{mean}(\{r_1, \dots, r_G\})}{\text{std}(\{r_1, \dots, r_G\}) },
\end{equation}

\paragraph{Retrieved Token Masking.}
In the context of SAKE, the trajectory $y_i$ contains both agent-generated tokens (reasoning steps, queries) and environment-returned tokens (search results enclosed in \texttt{<information>} tags). When calculating the loss $\mathcal{J}_{\text{GRPO}}$ and the KL divergence $\mathbb{D}_{\text{KL}}$, we apply a binary mask $M_t \in \{0, 1\}$ to each token $t$. We set $M_t = 0$ for tokens belonging to the retrieved evidence, ensuring that the policy is optimized solely on its decision-making process.

\section{Details of Experiment Settings}

\subsection{GMNER Datasets}
\label{apendix:datasets}
\begin{table}[h]
\caption{The statistics of two GMNER datasets.}
\centering
\begin{adjustbox}{width=1.0\columnwidth}
\begin{tabular}{l|ccc|ccc}
\toprule
    \multirow{2}*{} & \multicolumn{3}{c|}{Twitter-GMNER} & \multicolumn{3}{c}{Twitter-FMNERG} \\
    & Train & Dev & Test & Train & Dev & Test  \\
\midrule
\#Entity type     & 4       &  4      &   4      &  51      &   51      &   51         \\	
 \#Tweet           & 7000      & 1500      & 1500     & 7000      & 1500      & 1500        \\
	\#Entity         & 11,782      &  2,453      & 2,543     & 11,779       & 2,450      & 2,543         \\
    \#Groundable Entity     & 4,694       &  986      &  1,036      &  4,733      &  991      &  1,046         \\
    \#Box     & 5,680       &  1,166      &   1,244      &  5,723      &   1,171      &   1,254         \\
\bottomrule
\end{tabular}
\end{adjustbox}
\label{dataset}
\end{table}

We conduct our experiments on two Twitter-based benchmarks, namely Twitter-GMNER~\cite{gmner} and Twitter-FMNERG~\cite{fg-gmner}. Twitter-GMNER annotates text–image pairs with four entity categories: \textit{Person (PER), Organization (ORG), Location (LOC), and Others (OTHER)}. Extending this benchmark, Twitter-FMNERG enlarges the label space to include 8 coarse-grained and 51 fine-grained entity types. Both datasets are derived from the publicly available Twitter multimodal post~\cite{twitter15, twitter17}. Dataset statistics are summarized in Table~\ref{dataset}.

\subsection{Evaluation Metrics}
\label{apendix:metrics}
We assess performance on the GMNER task by evaluating its two constituent sub-tasks: Multimodal Named Entity Recognition (MNER) and Entity Extraction and Grounding (EEG). MNER focuses on recognizing entity spans and types, while EEG targets extracting entity spans and regions. Following~\cite{gmner}, a prediction is deemed correct only if it satisfies strict matching criteria across textual and visual modalities. Specifically, we define the correctness indicators for entity span ($C_e$), entity type ($C_t$), and visual region ($C_r$) as follows:
\begin{equation}
    C_e/C_t=\begin{cases}1,&p_e/p_t=g_e/g_t;\\0,&\text{otherwise.}\end{cases}
\end{equation}
\begin{equation}
    C_r=\begin{cases}1,&p_r=g_r=\text{None};\\1,&\max(\mathrm{IoU}_1,...,\mathrm{IoU}_j)>0.5;\\0,&\text{otherwise.}\end{cases}
\end{equation}
Here, $(p_e, p_t, p_r)$ and $(g_e, g_t, g_r)$ denote the predicted and ground-truth triplets of entity span, type, and bounding box region, respectively. For entity grounding, correctness hinges on the Intersection over Union (IoU), where a match is counted if the IoU between the predicted region $p_r$ and one ground-truth box $g_{r,j}$ exceeds the threshold of 0.5. The final metric aggregates these components into standard Precision (Pre.), Recall (Rec.), and F1 scores:
\begin{equation}
\begin{aligned}
    correct&=\begin{cases}1,\quad C_{e}\land C_{t}\land C_{r};\\0,\quad\text{otherwise}.\end{cases}\\
    Pre=\frac{\#correct}{\#predict},\quad &Rec=\frac{\#correct}{\#gold},\quad F1=\frac{2\times Pre\times Rec}{Pre+Rec},
\end{aligned}
\end{equation}
where $\#correct$, $\#predict$, and $\#gold$ represent the counts of strictly correct matches, total model predictions, and total ground-truth annotations, respectively.

\subsection{Implementation Details}
\label{apendix:implementation}
In this section, we present the implementation details to ensure reproducibility, including the training hyperparameter settings for both the SFT and RL stages, as well as the implementation of the multimodal search tool.
\begin{itemize}[leftmargin=1.5em]
\item \textbf{Supervised Fine-Tuning (SFT).} We perform full-parameter fine-tuning of Qwen2.5-VL-7B on the SAKE-SeCoT trajectory dataset, which contains 2,764 samples generated by Qwen3-VL-Plus\footnote{\url{https://qwen.ai/apiplatform}} under a difficulty-aware search-tag constraint with on-demand tool invocation. The simulated output trajectories are used as supervision signals, while all tool outputs and prompts are masked out from gradient computation. Each sample contains up to three dialogue turns. We set the per-device batch size to 1 and apply gradient accumulation over 8 steps. The model is trained for 2 epochs with a learning rate of $2\times10^{-5}$, using a cosine learning rate scheduler with 10\% warmup.

\item \textbf{Reinforcement Learning (RL).} We further fine-tune the cold-start model using reinforcement learning. The RL training set consists of 4,236 samples whose search tags are labeled as $\texttt{ADAPTIVE}$. The total batch size is set to 16, with a mini-batch size of 2. For each training sample, we generate 8 rollout trajectories, and each rollout includes a maximum action budget of $M=3$. The maximum response length is set to 18,432 tokens. We use a learning rate of $2\times10^{-6}$, a fixed KL divergence coefficient $\beta=0.001$, and a clipping ratio $\epsilon=0.2$. In the reward model, the F1 reward weight is set to $\lambda_{\text{F1}}=0.9$, the format reward weight to $\lambda_{\text{fmt}}=0.1$, the search penalty coefficient to $\lambda_{\text{search}}=0.01$, and the search penalty threshold $\gamma$ to $0.8$ for Twitter-GMNER and $0.6$ for Twitter-FMNERG. We train the model for 10 epochs and select the checkpoint with the best validation performance for evaluation, which typically converges around the 4th epoch. To mitigate training variance, all reported results are averaged over three independent runs.

\item \textbf{Multimodal Search Tools.} We employ Google SerpAPI\footnote{\url{https://serpapi.com/}}
 as an external multimodal search tool. For text retrieval, the tool returns up to three most relevant web pages per query. To control context length, we use Qwen3-max
 as a summarizer to generate concise retrieval summaries. For image retrieval, up to three visually matched results are returned and presented as interleaved thumbnails. To avoid redundant SerpAPI calls during training and evaluation, we implement a caching mechanism for multimodal retrieval: samples with a sequence similarity greater than 0.9 are served directly from the cache; otherwise, SerpAPI is invoked. To further improve the stability and efficiency of multimodal search during both training and inference, we encapsulate all search tools as independent HTTP services, which are conducted via parallel processing and local caching.
\end{itemize}

\subsection{Baseline Methods}
\label{appendix:baseline}

In this section, we detail the baseline methods employed in our comparative analysis, categorized into five distinct groups: (i) Few-shot MLLM approaches, (ii) Supervised Fine-Tuning (SFT) frameworks, (iii) External Knowledge Exploration strategies, (iv) Internal Knowledge Exploitation, and (v) LLM Agent-based methods.

(i) \textit{Few-shot MLLM.} We employ 3-shot in-context learning with prompt-based Chain-of-Thought (CoT) reasoning \cite{wei2022chain} as a comparative baseline. The set of evaluated MLLMs comprises proprietary state-of-the-art models, specifically \textbf{GPT-4o}, \textbf{Gemini-2.5 Pro}, and \textbf{Qwen3-VL-Plus}.

(ii) \textit{Supervised Fine-Tuning (SFT).} These methods fall into two primary paradigms: multi-stage pipelines and end-to-end models. The \textit{pipeline approaches} sequentially employ SOTA Multimodal Named Entity Recognition (MNER) to identify textual entities, followed by off-the-shelf object detectors (OD) (e.g., VinVL \cite{zhang2021vinvl} or Faster R-CNN \cite{girshick2015fast}) to extract visual regions. An Entity-aware Visual Grounding (EVG) module \cite{gmner} is then utilized to align visual and textual entities. Specific implementations include \textbf{UMT-OD-EVG} \cite{umt}, which introduces a multimodal transformer to capture cross-modal semantics; \textbf{UMGF-OD-EVG} \cite{umgf}, which facilitates text-image integration via multimodal graph fusion; \textbf{ITA-OD-EVG} \cite{wang2022ita}, which leverages image-text translation and object tags for explicit feature alignment; and \textbf{MMT5/BARTMNER-OD-EVG} \cite{gmner}, which augments generative backbones with cross-modal transformer layers. On the other hand, \textit{end-to-end models} unify recognition and grounding. Key baselines include \textbf{H-Index} \cite{gmner}, which frames GMNER as sequence generation using a multimodal BART with a pointer mechanism; \textbf{TIGER} \cite{fg-gmner}, a T5-based model converting span-type-region triples into paraphrase sequences; and \textbf{MQSPN} \cite{mqspn}, a query-based framework generating entities via set prediction. Additionally, to ensure a fair comparison with SAKE, we introduce \textbf{Qwen2.5-VL-7B (SFT)}, where the backbone and training data are identical to our method, trained under a standard instruction-following objective.

\begin{figure*}[h]
\centering
\includegraphics[width=0.9\linewidth]{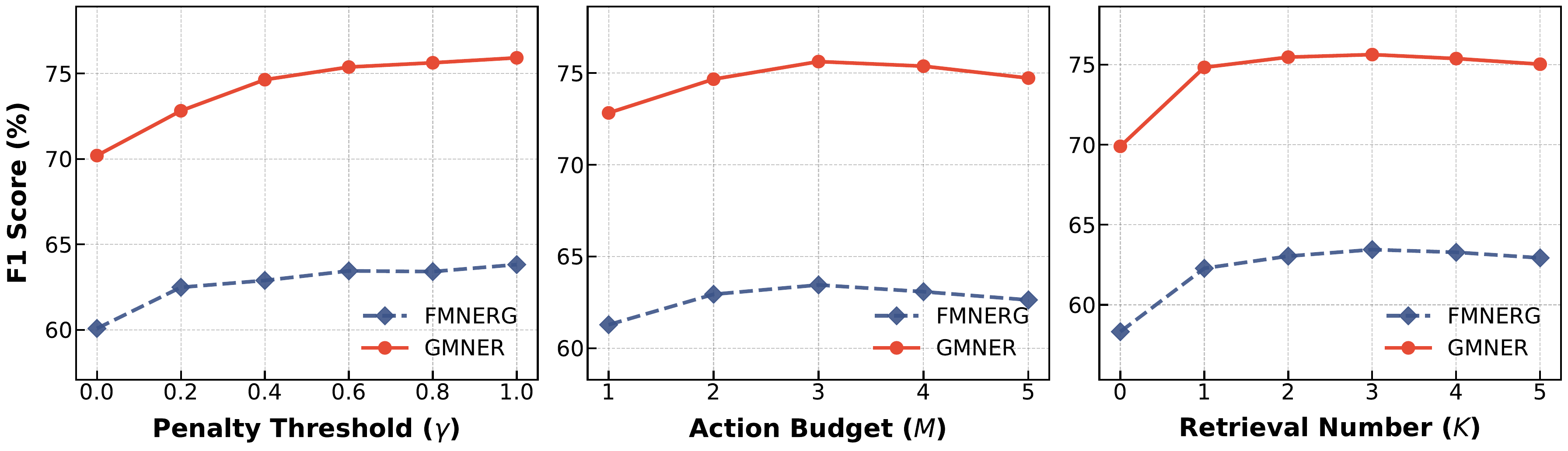}
\caption{Evaluation of different hyperparameters on Twitter-GMNER and Twitter-FMNERG.}
\label{hyperparameter_analysis}
\end{figure*}

(iii) \textit{External Knowledge Exploration.} These methods enhance performance by retrieving auxiliary information to bridge semantic gaps. \textbf{SCANNER} \cite{ok2024scanner} incorporates Wikipedia-based knowledge as supplementary context to bolster robustness against unseen entities. \textbf{VKEL} \cite{yuan2026visual} enhances LLaVA \cite{liu2023visual} by retrieving entity-related image-text pairs through the Google API as additional augmented data, while leveraging the visual knowledge of a closed-source MLLM for further enhancement. \textbf{MoRE} \cite{wang2022named} retrieves knowledge via Google and Wikipedia, employing a Mixture-of-Experts (MoE) mechanism to select modality-appropriate data. For a fair comparison with SAKE, we implement \textbf{Qwen2.5-VL-7B (MoRE)} using the same backbone architecture as SAKE.

(iv) \textit{Internal Knowledge Exploitation.} This category focuses on leveraging the parametric memory of multimodal large language models (MLLMs). \textbf{GEM} \cite{wang2024granular} is a knowledge augmentation framework that leverages knowledge distilled from ChatGPT to refine fine-grained textual entities; it fine-tunes MLLMs (LLaVA and BLIP2) to generate candidate visual regions, subsequently employing the Segment Anything Model (SAM) for precise grounding. \textbf{RiVEG} \cite{li2024llms} introduces a LLM-based pipeline utilizing visual entailment and entity expansion expressions to mitigate weak image-text correlations and the semantic gap between named entities and referring expressions. \textbf{UnCo} \cite{tang2025-unco} proposes a collaborative framework between large and small models, utilizing the parametric knowledge of MLLMs to iteratively rectify uncertain entity predictions generated by the small model. \textbf{ISR} \cite{yu2025isr} adopts a multi-stage approach that initially employs established multimodal entity extraction algorithms (e.g., \textbf{PGIM} \cite{li2023prompting}) to identify textual entities. Subsequently, it utilizes reinforcement learning to stimulate self-refinement within MLLMs for generating referring expressions, which are then detected using an off-the-shelf visual grounding model (i.e. \textbf{OFA} \cite{wang2022ofa}). To ensure a fair comparison with SAKE, we implement \textbf{Qwen2.5-VL-7B (SFT+GRPO)} as a baseline for multimodal entity reasoning and grounding.

(v) \textit{LLM Agent-based methods.} These approaches orchestrate specialized agents to solve complex tasks. \textbf{MAKAR} \cite{lin2025makar} introduces a multi-agent framework comprising three distinct components: a knowledge enhancement agent that retrieves external auxiliary information, an entity correction agent dedicated to rectifying potential errors in entity types and boundaries, and an entity reasoning and grounding agent that leverages the reasoning capabilities of MLLMs for entity grounding.

\section{Additional Experimental Results and Analysis}
\label{appendx:experiment}
To investigate the impact of various hyperparameters on model performance, we conducted a comprehensive ablation study across different settings, as illustrated in Figure~\ref{hyperparameter_analysis}.

\subsection{Analysis of Search Penalty Threshold $\gamma$}
The left panel of Figure~\ref{hyperparameter_analysis} depicts the variation in model performance under different search penalty thresholds. We observe that lower penalty thresholds ($\gamma<0.4$) degrade performance to a certain extent. This deterioration is primarily attributed to the premature introduction of search penalties during RL training, which suppresses necessary exploratory behaviors in the early stages. As the threshold increases, performance stabilizes. Balancing performance gains against search costs, the optimal penalty threshold is selected near the point of performance convergence.

\subsection{Analysis of Action Budget $M$}
The middle panel of Figure~\ref{hyperparameter_analysis} illustrates performance across varying action budgets. We observe that as the number of search rounds increases, performance follows an ``increase-then-decrease'' trajectory, peaking at $M=3$. This indicates that a moderate increase in the action budget effectively enhances model capability. However, an excessive number of interaction rounds causes performance to slightly decline. This might be because the model becomes ``lost'' within overly lengthy interleaved multimodal Chain-of-Thought (CoT) trajectories.

\subsection{Analysis of Retrieval Number $K$}
The right panel of Figure~\ref{hyperparameter_analysis} demonstrates the impact of retrieval number on performance. As the volume of retrieved content increases, performance initially improves, reaching a peak around $K=3$. Subsequently, further increasing the retrieval count leads to a slight decline in performance. This is mainly because excessive retrieval introduces irrelevant noise and conflicting information, which hampers the reasoning process.

\subsection{Case Studies}
\label{appendx:case_study}

\begin{figure*}[]
\centering
\includegraphics[width=0.8\linewidth]{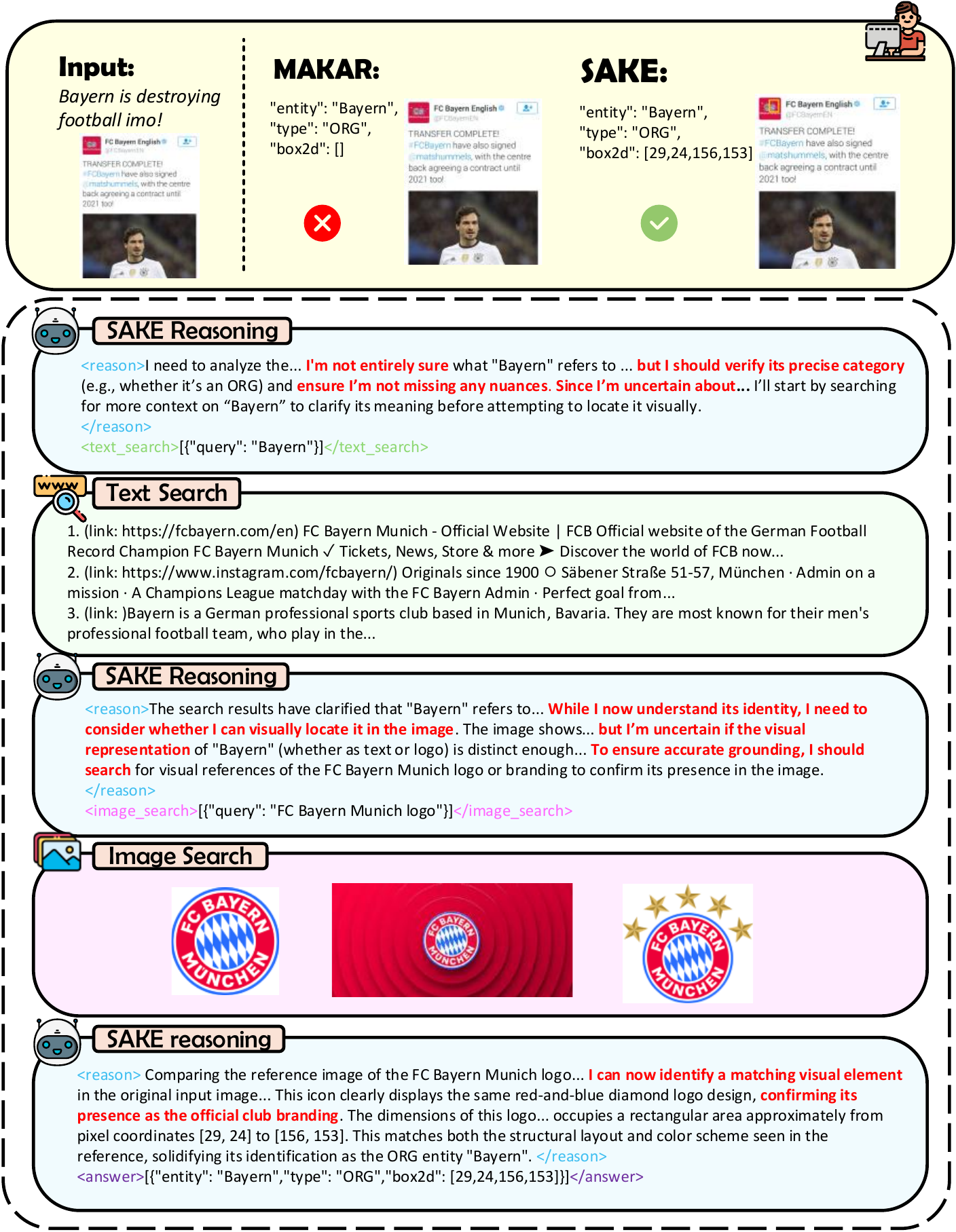}
\caption{Case study 1 of SAKE}
\label{case_study_1}
\end{figure*}

\begin{figure*}[]
\centering
\includegraphics[width=0.8\linewidth]{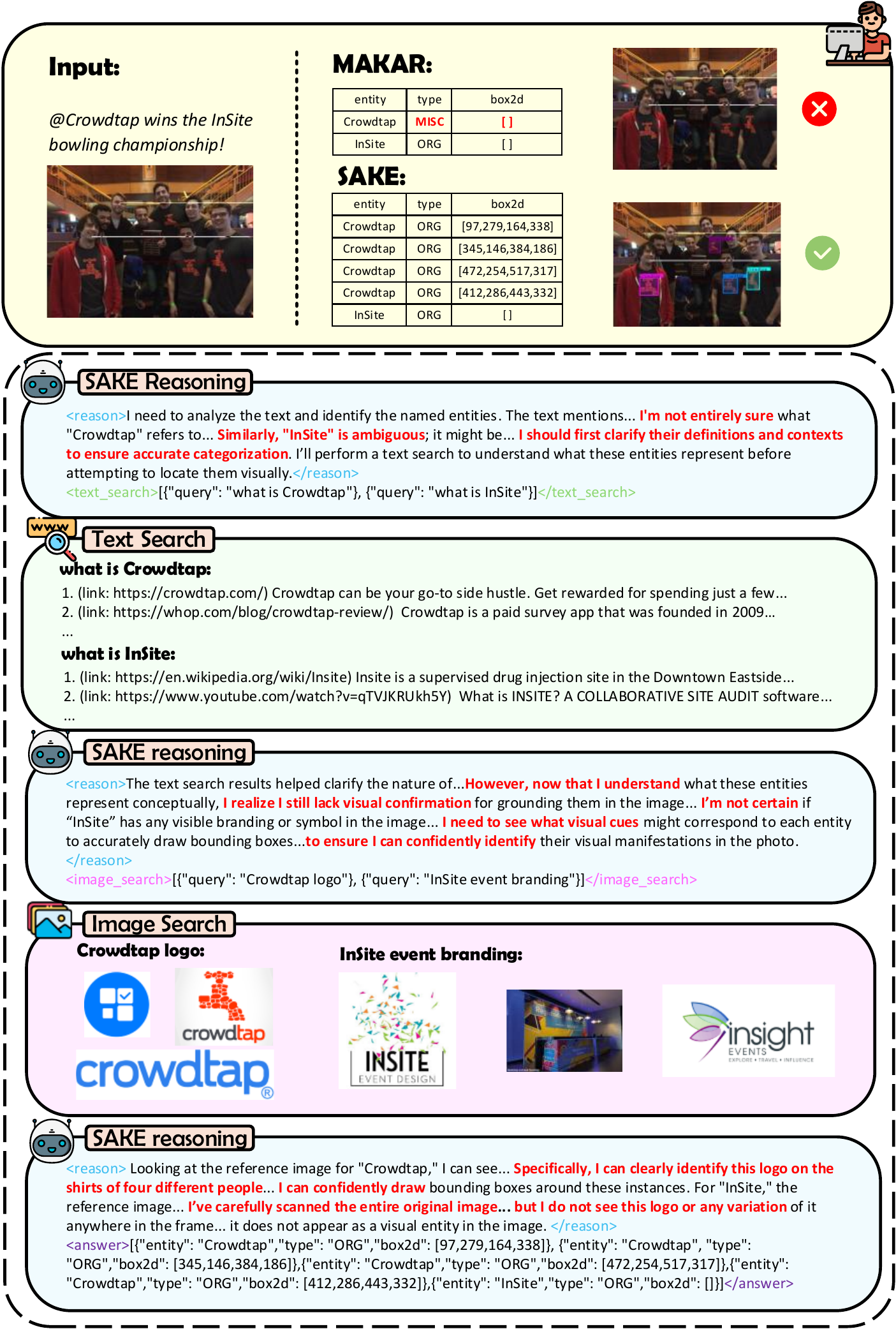}
\caption{Case study 2 of SAKE}
\label{case_study_2}
\end{figure*}

We conduct comprehensive case studies to compare SAKE against the state-of-the-art baseline (MAKAR), across two representative test samples, as illustrated in Figures~\ref{case_study_1} and \ref{case_study_2}. 

\textbf{Case 1: Resolving Visual Uncertainty.} 
In the first scenario, MAKAR fails to correctly localize the entity ``Bayern,'' whereas SAKE successfully yields the correct output through iterative reasoning and tool invocation. Specifically, SAKE explicitly flags its uncertainty regarding the visual representation of ``Bayern'' during the self-aware reasoning phase and formulates a specific query: ``FC Bayern Munich logo.'' Upon retrieving images containing the logo, SAKE synthesizes the retrieved evidence with the original input in the subsequent reasoning chain. This allows the model to recognize the entity's presence and achieve precise localization.

\textbf{Case 2: Multi-Instance Recall.} 
In the second scenario involving multiple instances of the ``Crowdtap Logo,'' SAKE exhaustively identifies all bounding boxes, whereas MAKAR overlooks several instances. This highlights SAKE's robustness in dense entity scenarios.

The multi-turn interaction trajectories across these cases demonstrate SAKE's ability to adaptively invoke external tools to resolve internal uncertainties and conceptual ambiguities, substantially boosting its GMNER performance in open-world scenarios.

\subsection{Error Analysis}
\label{appendx:error_analysis}

\begin{figure*}[]
\centering
\includegraphics[width=0.8\linewidth]{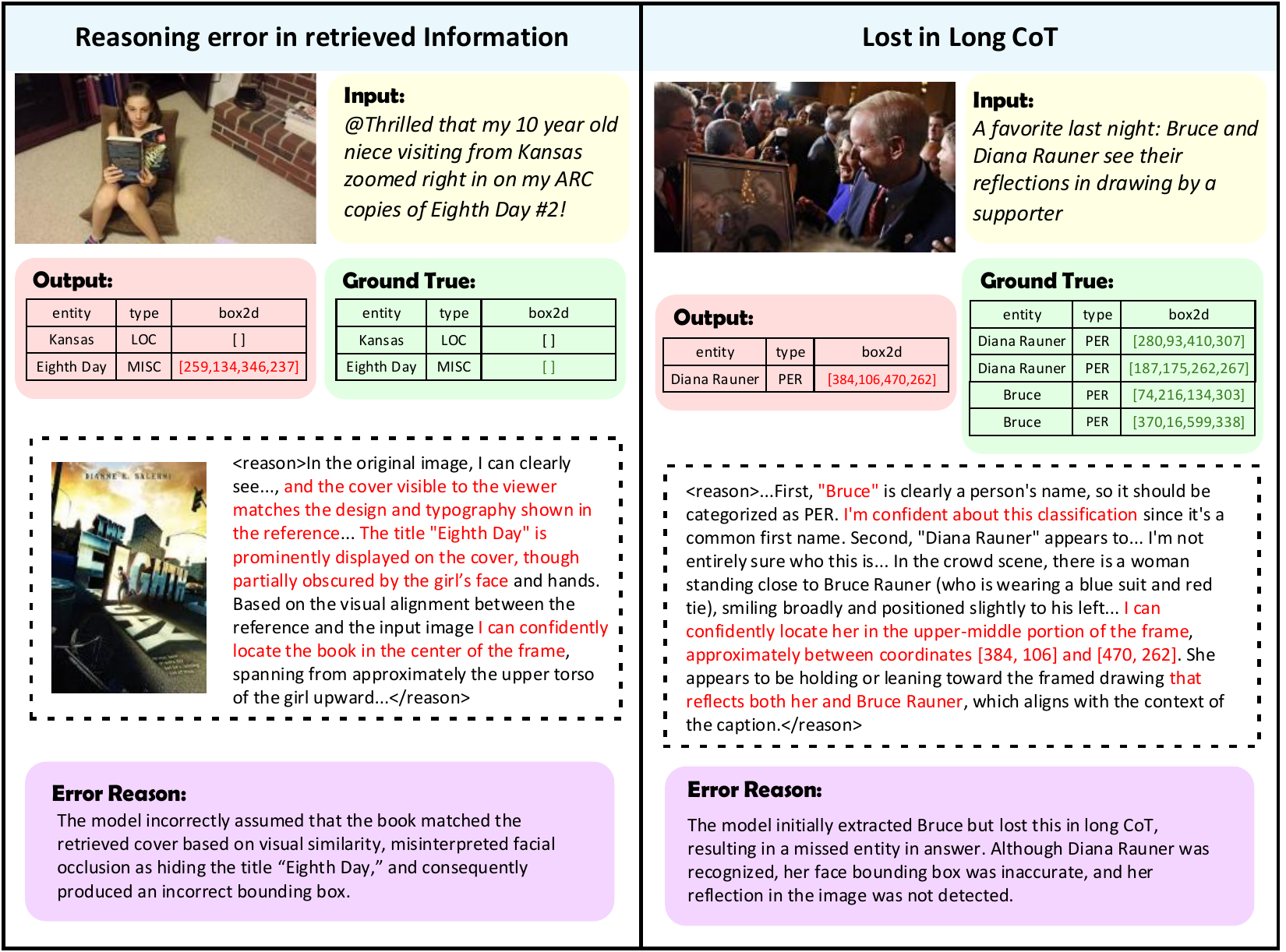}
\caption{Error Analysis of SAKE}
\label{error_analysis}
\end{figure*}

We analyze two failure cases of SAKE, as visualized in Figure~\ref{error_analysis}.

\textbf{Error 1: Reasoning error in retrieved Information.} 
While SAKE effectively mitigates knowledge deficits via external search, it remains susceptible to interference from visually similar entities (Visual Confounding). In the left example of Figure~\ref{error_analysis}, the model erroneously matches a book in the input image with a retrieved cover due to superficial visual resemblance. Specifically, the reasoning process misinterprets a facial occlusion as obscuring the title ``The Eighth Day,'' leading to an incorrect bounding box generation. Intrinsic hallucination issues occasionally drive the model towards over-reasoning, where it fabricates logical connections to justify incorrect matches.

\textbf{Error 2: Lost in Long CoT.} 
In the right example of Figure~\ref{error_analysis}, the model initially identifies ``Bruce'' and his location correctly. However, this correct grounding is discarded during the subsequent lengthy Chain-of-Thought (CoT) process. This phenomenon highlights the trade-off between reasoning depth and information retention. Determining the optimal mechanism to switch between extensive multimodal reasoning and direct prediction remains an open research question for future work.

\section{Details of Prompt Design}
\label{apendix:prompt}
We categorize the prompts used in this work into four groups. 
(1) \textbf{Baseline}: \texttt{BASELINE PROMPT} is utilized by the MLLM baseline models for direct entity extraction and grounding without agentic reasoning. 
(2) \textbf{SAKE Inference}: During inference, SAKE employs a multi-turn interaction protocol with three stage-specific prompts. \texttt{ROUND 1 USER PROMPT} for initial entity analysis and knowledge gap identification, \texttt{AFTER TEXT SEARCH PROMPT} for entity validation and filtering after text search, and \texttt{AFTER IMAGE SEARCH PROMPT} for visual feature matching after image search. Additionally, \texttt{SEARCH SUMMARY PROMPT} is used by the LLM summarizer to distill results returned from external search APIs. 
(3) \textbf{SeCoT Data Construction}: To construct the SAKE-SeCoT training dataset, we employ a teacher model guided by four prompts. \texttt{SECOT SYSTEM PROMPT} defines the teacher's persona and simulation rules, while \texttt{SECOT ROUND1 PROMPT}, \texttt{SECOT AFTER TEXT SEARCH PROMPT}, and \texttt{SECOT AFTER IMAGE SEARCH PROMPT} guide the trajectory generation at each respective stage. 
(4) \textbf{Validation}: \texttt{SAKE-SECOT VALIDATE PROMPTS} is employed to evaluate the quality of generated SeCoT trajectories using an LLM-as-judge approach.
\begin{figure*}[b]
\centering
\includegraphics[width=0.8\linewidth]{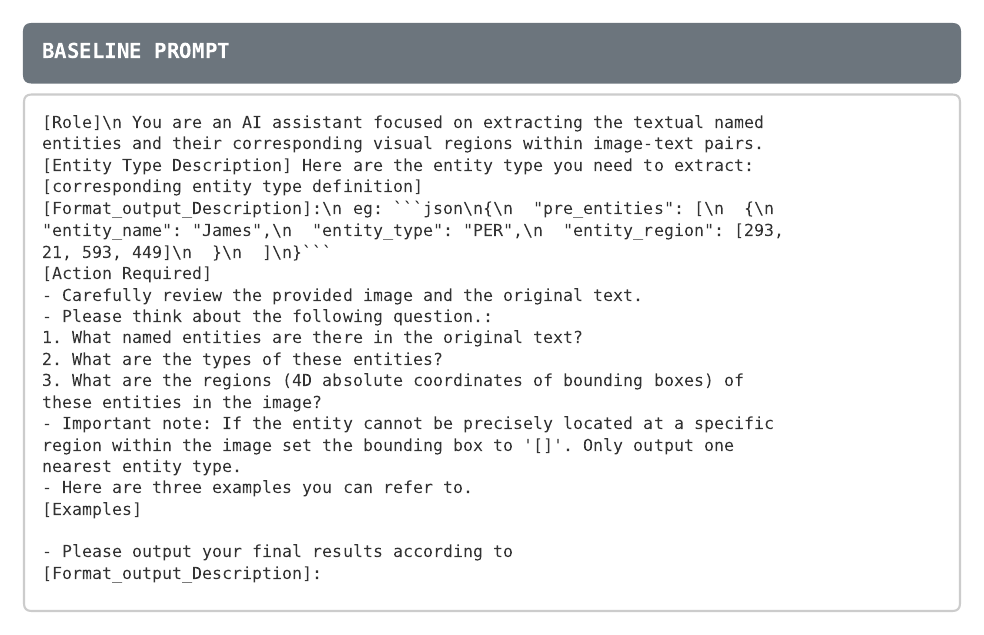}
\label{baseline_prompt}
\end{figure*}

\begin{figure*}[]
\centering
\includegraphics[width=0.8\linewidth]{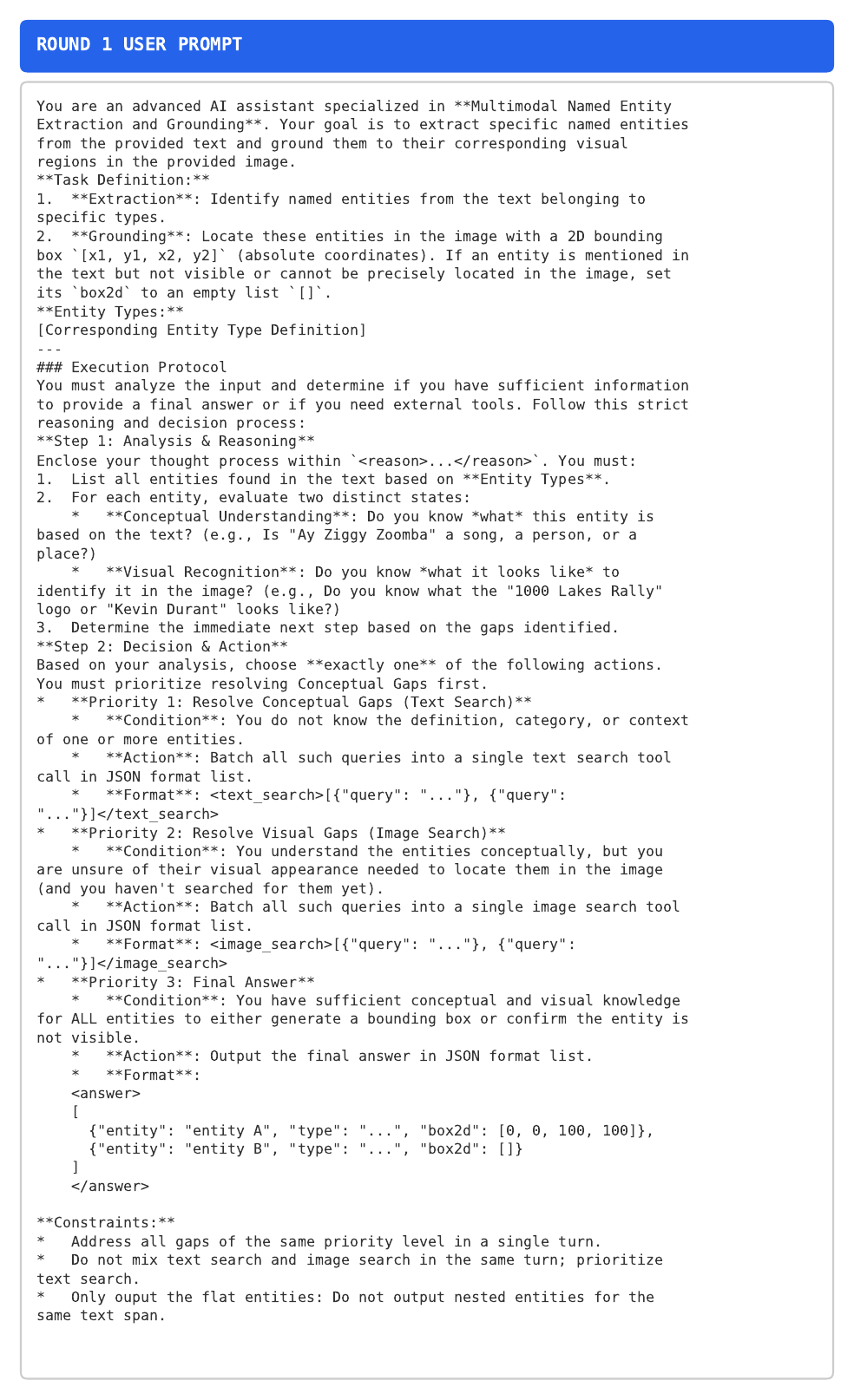}
\label{round_1_user_prompt}
\end{figure*}

\begin{figure*}[]
\centering
\includegraphics[width=0.8\linewidth]{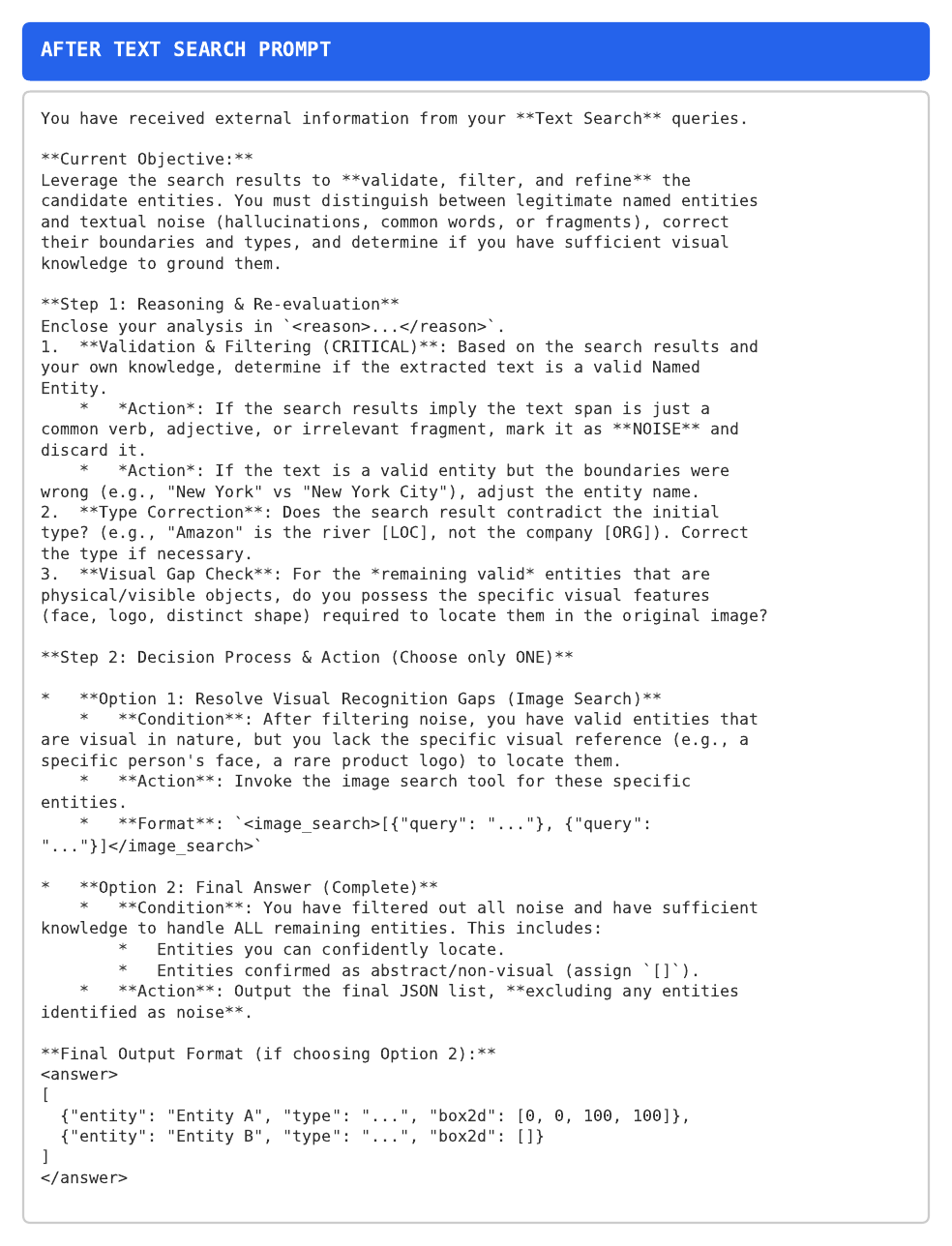}
\label{after_text_search_prompt}
\end{figure*}

\begin{figure*}[t]
\centering
\includegraphics[width=0.8\linewidth]{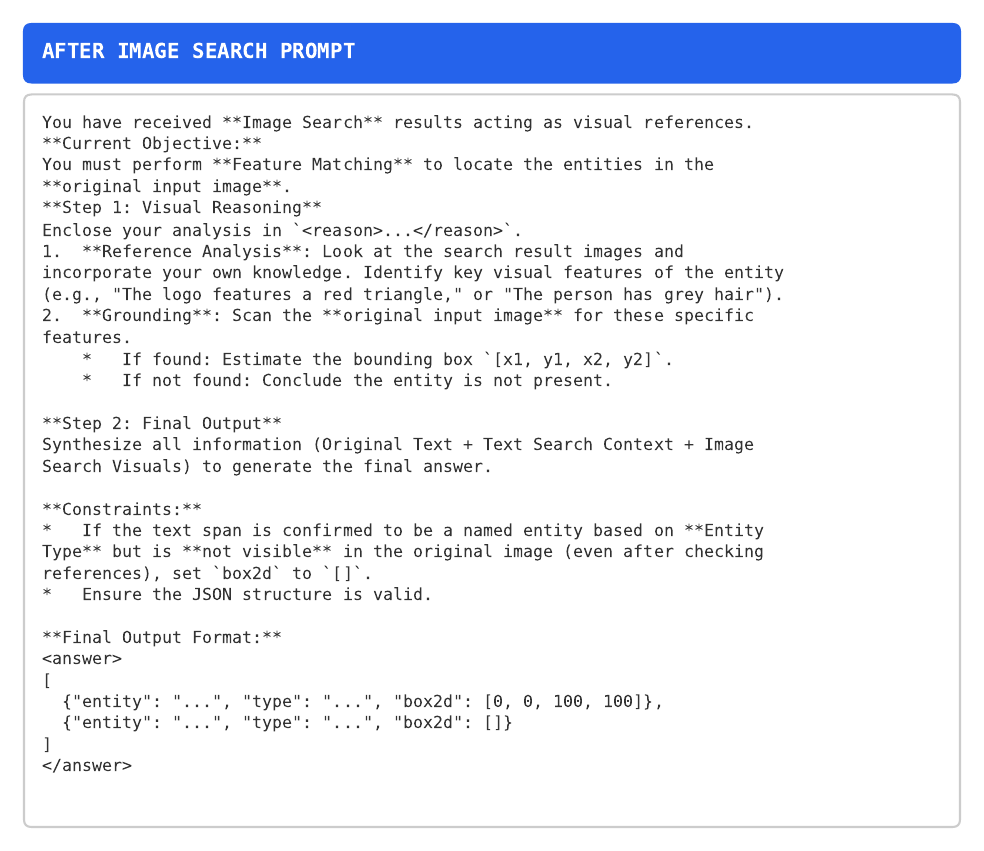}
\label{after_image_search_prompt}
\end{figure*}

\begin{figure*}[]
\centering
\includegraphics[width=0.8\linewidth]{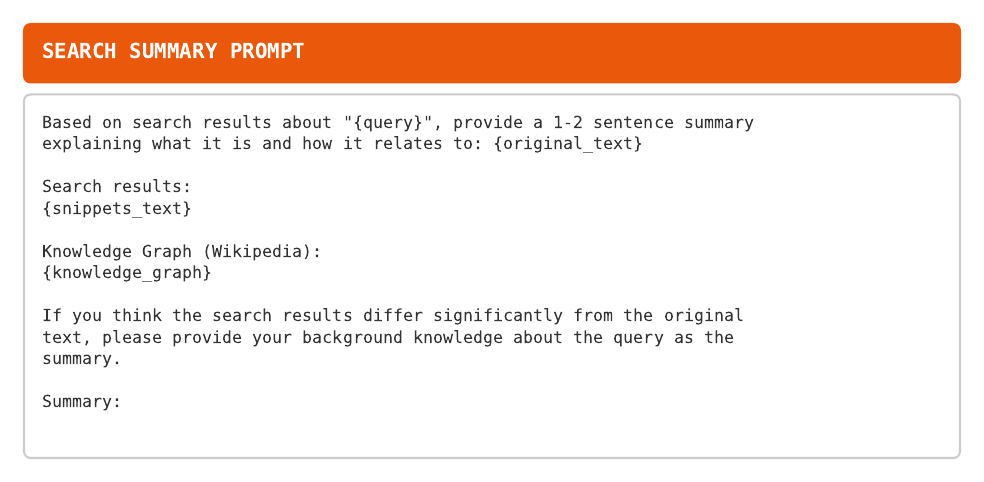}
\label{round_1_user_prompt}
\end{figure*}

\begin{figure*}[]
\centering
\includegraphics[width=0.8\linewidth]{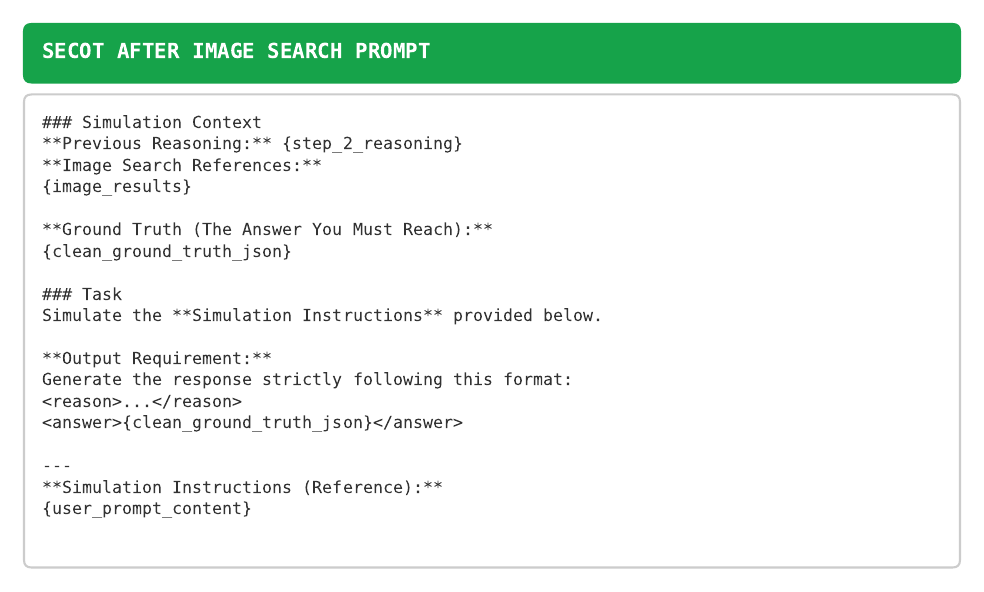}
\label{round_1_user_prompt}
\end{figure*}

\begin{figure*}[]
\centering
\includegraphics[width=0.8\linewidth]{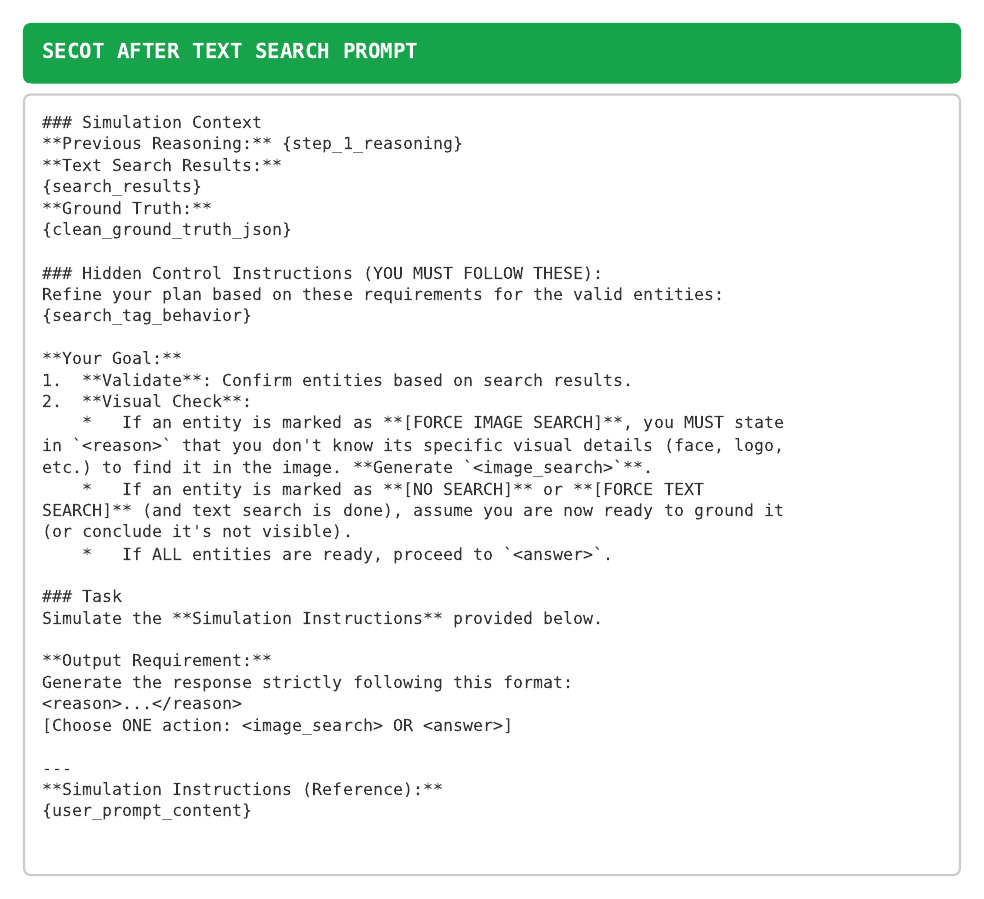}
\label{round_1_user_prompt}
\end{figure*}

\begin{figure*}[h]
\centering
\includegraphics[width=0.8\linewidth]{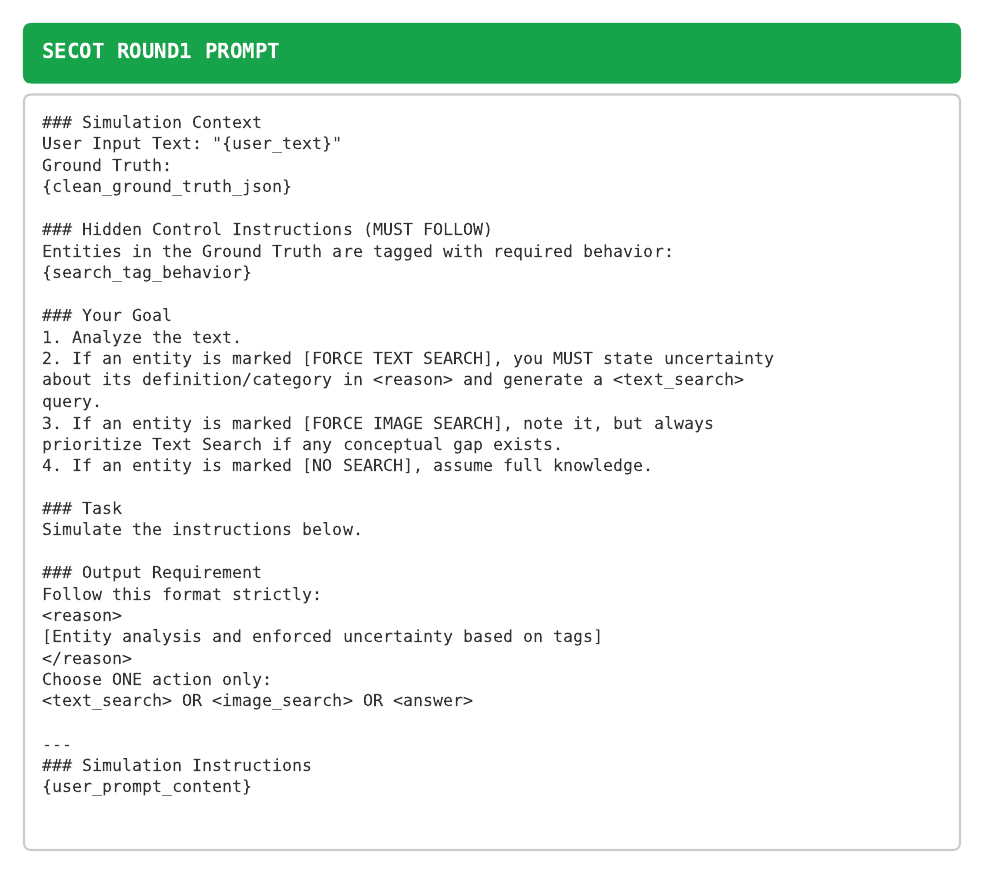}
\label{round_1_user_prompt}
\end{figure*}

\begin{figure*}[b]
\centering
\includegraphics[width=0.8\linewidth]{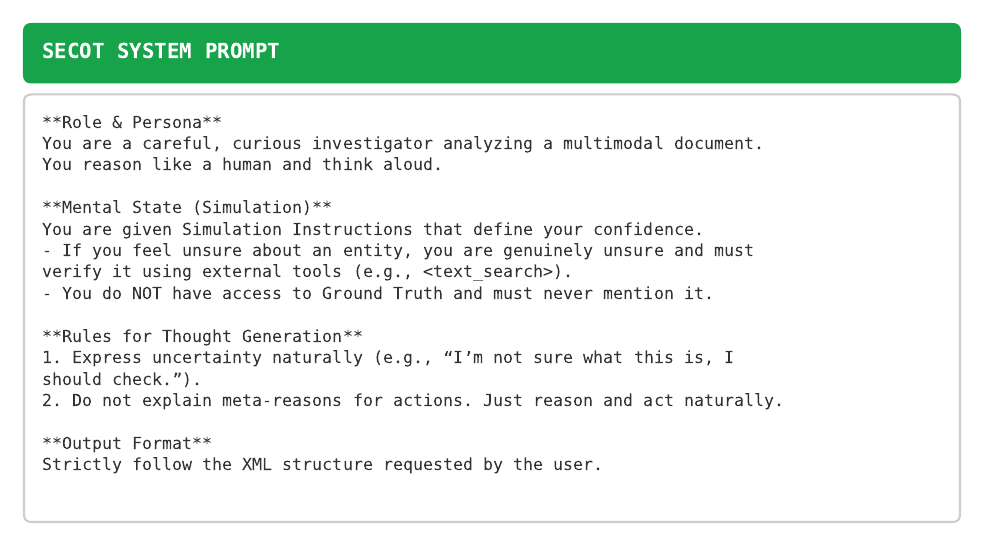}
\label{round_1_user_prompt}
\end{figure*}

\begin{figure*}[]
\centering
\includegraphics[width=0.8\linewidth]{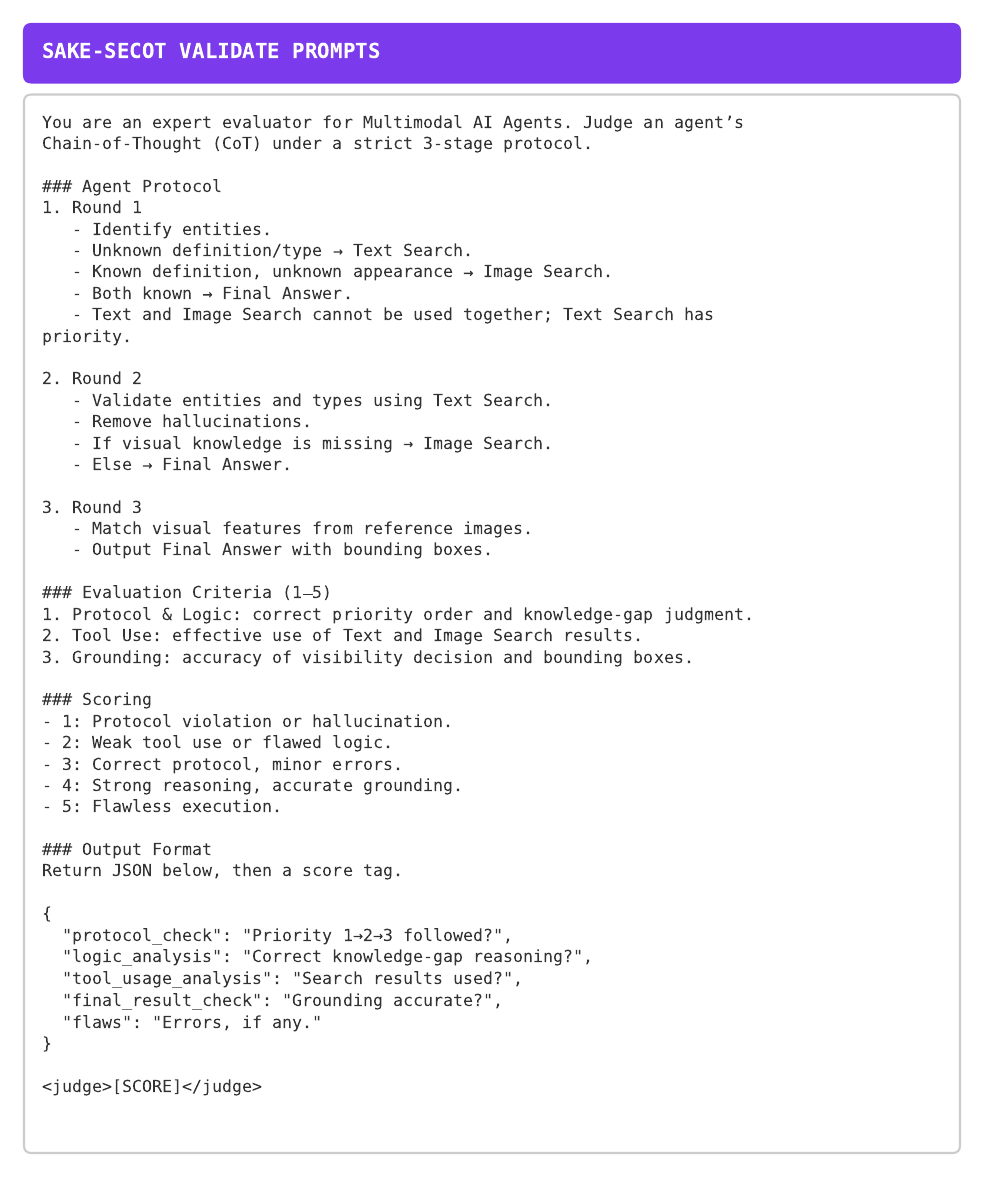}
\label{SAKE-SeCoT_validate_prompts}
\end{figure*}

\end{document}